\documentclass{aa}
\usepackage[varg]{txfonts}

\usepackage{natbib}
\bibpunct{(}{)}{;}{a}{}{,} 
\usepackage{array}

\usepackage{newtxtext,newtxmath}

\usepackage[T1]{fontenc}
\usepackage{ae,aecompl}

\usepackage{graphicx}	
\usepackage{amsmath}	
\usepackage{amssymb}	
\usepackage{lscape}
\usepackage{comment}    
\usepackage{ulem}           
\usepackage{adjustbox}

\begin{document}

\title{The XXL Survey}
\subtitle{XVIII. ATCA 2.1 GHz radio source catalogue and source counts for the XXL-South field}


\author{Andrew Butler\inst{\ref{inst0}}\thanks{E-mail: andrew.butler@icrar.org} \and Minh Huynh\inst{\ref{inst0},\ref{inst1}} \and Jacinta Delhaize\inst{\ref{inst2}} \and Vernesa Smol\v{c}i\'{c}\inst{\ref{inst2}} \and Anna Kapi\'{n}ska\inst{\ref{inst0}} \and Dinko Milakovi\'{c}\inst{\ref{inst2}} \and Mladen Novak\inst{\ref{inst2}} \and Nikola Baran\inst{\ref{inst2}} \and Andrew O'Brien\inst{\ref{inst3}} \and Lucio Chiappetti\inst{\ref{inst4}} \and Shantanu Desai\inst{\ref{inst5}} \and Sotiria Fotopoulou\inst{\ref{inst6}} \and Cathy Horellou\inst{\ref{inst7}} \and Chris Lidman\inst{\ref{inst8}} \and Marguerite Pierre\inst{\ref{inst9}}}

\institute{International Centre for Radio Astronomy Research (ICRAR), University of Western Australia, 35 Stirling Hwy, Crawley WA 6009, Australia\label{inst0}
\and
CSIRO Astronomy and Space Science, 26 Dick Perry Ave, Kensington WA 6151, Australia\label{inst1}
\and
Physics Department, University of Zagreb, Bijeni\v{c}ka cesta 32, 10002 Zagreb, Croatia\label{inst2}
\and
Western Sydney University, Locked Bag 1797, Penrith NSW 2751, Australia\label{inst3}
\and
INAF, IASF Milano, via Bassini 15, I-20133 Milano, Italy\label{inst4}
\and
Department of Physics, IIT Hyderabad, Kandi, Telangana-502285, India\label{inst5}
\and
Department of Astronomy, University of Geneva, Ch. d'\'{E}cogia 16, 1290 Versoix, Switzerland\label{inst6}
\and
Onsala Space Observatory, Department of Earth and Space Sciences, Chalmers University of Technology, SE-439 92 Onsala, Sweden\label{inst7}
\and
Australian Astronomical Observatory, 105 Delhi Rd, North Ryde NSW 2113, Australia\label{inst8}
\and
Service d'Astrophysique AIM, CEA/DRF/IRFU/SAp, CEA Saclay, 91191 Gif-sur-Yvette, France\label{inst9}}

\date{Received 25 November 2016 / Accepted 12 March 2017}

\abstract{The 2.1 GHz radio source catalogue of the 25 deg$^2$ ultimate XMM extragalactic survey south (XXL-S) field, observed with the Australia Telescope Compact Array (ATCA), is presented.  The final radio mosaic achieved a resolution of $\sim$$4.8''$ and a median rms noise of $\sigma \approx41$ $\mu$Jy/beam.  To date, this is the largest area radio survey to reach this flux density level.  A total of 6350 radio components above 5$\sigma$ are included in the component catalogue, 26.4\% of which are resolved.  Of these components, 111 were merged together to create 48 multiple-component radio sources, resulting in a total of 6287 radio sources in the source catalogue, 25.9\% of which were resolved.  A survival analysis revealed that the median spectral index of the Sydney University Molonglo Sky Survey (SUMSS) 843 MHz sources in the field is $\alpha$ = $-$0.75, consistent with the values of $-0.7$ to $-0.8$ commonly used to characterise radio spectral energy distributions of active galactic nuclei (AGN).  The 2.1 GHz and 1.4 GHz differential radio source counts are presented and compared to other 1.4 GHz radio surveys.  The XXL-S source counts show good agreement with the other surveys.}





\authorrunning{Butler et al.}
\maketitle





\section{Introduction}

Star formation and active galactic nuclei (AGN) activity are two important processes that influence galaxy evolution.  AGN in particular have been recognised as having a major influence on massive galaxy evolution via a process called feedback (e.g. \citealp{bohringer1993,forman2005,fabian2012}). Incorporating this feedback into galaxy evolution models gives a markedly better fit to the optical luminosity function for galaxies at $z \leq 0.2$, particularly at the high-mass end \citep{croton2006, bower2006}.

Studies of galaxies at radio frequencies of $\sim$2 GHz probe these processes via the non-thermal synchrotron emission generated by relativistic electrons spiraling around magnetic field lines (e.g. \citealp{condon1992}).  Since this emission is not obscured by dust and is detectable at large cosmic distances, radio surveys provide a less dust-biased picture of galaxy evolution than optical surveys (e.g. \citealp{haarsma2000} and \citealp{seymour2008}).  They also allow the construction of the radio luminosity function (RLF), which is the most direct and accurate way to measure the cosmic evolution of radio sources \citep{sadler2007}.  This is essential for measuring the relative contribution of the star-forming galaxy and AGN populations to the total radio power emitted at a given epoch \citep{mauch2007}.  Measuring how the shapes of the AGN and star-forming galaxy RLFs evolve over cosmic time provides important constraints on galaxy evolution models.

A number of studies have used the RLF to determine the cosmic evolution of the radio AGN population across a wide range of radio luminosity.  For example, \cite{dunlop1990} and \cite{rigby2011} found that high-luminosity radio AGN ($L_{\rm{1.4GHz}}\gtrsim~10^{26}$ W/Hz) increase their comoving density by a factor of $\sim$$100-1000$ out to $z \sim 2-3$.  On the other hand, \cite{smolcic2009} investigated the evolution of the low-luminosity radio AGN ($L_{\rm{1.4GHz}} < 5 \times 10^{25}$ W/Hz) using VLA-COSMOS data and found an increase in density by a factor of $\sim$2 out to $z = 1.3$.  \cite{waddington2001}, \cite{mcalpine2011}, and \cite{williams2015} came to similar conclusions regarding the evolutionary differences between high- and low-luminosity radio AGN.

These results suggest that high- and low-luminosity radio AGN form two distinct populations.  Observations of the host galaxies of radio AGN have supported this idea, favouring a characterisation in which the high-luminosity radio AGN exhibit strong high-excitation emission lines (such as [\ion{O}{III}]) and the low-luminosity radio AGN do not (e.g. \citealp{hine1979} and \citealp{laing1994}).  The former are called high-excitation radio galaxies (HERGs) and the latter low-excitation radio galaxies (LERGs).  It is hypothesised that HERGs and LERGs exhibit fundamentally different black hole accretion modes that result in two distinct forms of feedback (e.g. \citealp{hardcastle2007}).

However, there have been few studies that separate HERG and LERG RLFs.  \cite{best2012} and \cite{best2014} constructed separate RLFs for the two populations down to 5 mJy out to $z\sim0.3$ and $0.5<z<1.0$, respectively.  \cite{pracy2016} probed down to 2.8 mJy out to $z\sim0.75$ and found that the LERG RLF is consistent with no evolution while the HERG RLF indicates rapid evolution.  The local LERG RLFs from these studies were consistent with each other, but the local HERG RLF from \cite{pracy2016} displayed higher space densities than that of \cite{best2012}, especially for $L_{\rm{1.4GHz}}\leq10^{24}$ W/Hz.  The probable reason for this discrepancy is that \cite{best2012} classified certain sources as star-forming galaxies that \cite{pracy2016} classified as HERGs.  Clearly, more data are needed to clarify this discrepancy and to gain a better understanding of the HERG and LERG luminosity functions, host galaxies, and cosmic evolution.  This requires a deep radio survey over a relatively wide area combined with excellent multiwavelength data in order to capture the largest possible range of radio luminosities out to $z\sim1$ \citep{sadler2007}.

To this end, a pilot 2.1 GHz radio survey covering the central $\sim$5 deg$^2$ of the XXL-South field was conducted with the Australia Telescope Compact Array (ATCA).  The ultimate XMM extragalactic (XXL) survey (\citealp{pierre2016_xxl1}, XXL Paper I) is the largest survey undertaken with the \textit{XMM-Newton} X-ray telescope.  The survey observed two 25 deg$^2$ fields, the XMM-LSS field (XXL-North, at $\alpha=2^{\text{h}}20^{\text{m}}00^{\text{s}}$, $\delta=-5^{\circ}00'00''$) and the BCS-XMM field (XXL-South, at $\alpha=23^{\text{h}}30^{\text{m}}00^{\text{s}}$, $\delta=-55^{\circ}00'00''$), over a total of 6.9 Ms.  One of the main goals of the XXL project is to provide a lasting legacy for studies of AGN, their nature, demographics, and evolution across cosmic time. The pilot ATCA observations of XXL-South achieved an average rms sensitivity of $\sim$50 $\mu$Jy/beam and an angular resolution of $\sim$4.4", confirming that the ATCA could reach the sensitivity and resolution required to construct evolving RLFs for the HERGs and LERGs in the field (\citealp{smolcic2016_pilot}, hereafter XXL Paper XI).

This paper describes the new ATCA radio observations of the remaining $\sim$20 deg$^2$ of the XXL-South field (hereafter XXL-S) and the analysis of the full $\sim$25 deg$^2$.  In Section~\ref{sec:obs_cal_img}, the observations, calibration, and imaging of XXL-S are described.  In Section~\ref{sec:mosaic_properties}, the properties of the final mosaic are presented.  Section~\ref{sec:final_cats} discusses the final radio component and source catalogues constructed from the mosaic.  The spectral indices of the sources and the radio source counts are presented in Sections \ref{sec:spectral_indices} and \ref{sec:source_counts}, respectively.  Section~\ref{sec:conclusions} contains the summary.

\section{Observations, calibration and imaging}
\label{sec:obs_cal_img}

\subsection{Pilot observations of XXL-S}

The ATCA pilot observations of XXL-S were performed in 2012 using 2048 1-MHz-wide channels of the Compact Array Broadband Backend (CABB) correlator \citep{wilson2011} from 1.076-3.124 GHz.  Table~\ref{tab:observations} shows the dates, configurations, and net observing times of all the observations.  The central $\sim$6.5 deg$^2$ of XXL-S were observed, which required 81 pointings arranged in a square pattern.  The survey detected 1389 radio sources above 5$\sigma$ within the inner $\sim$5 deg$^2$ masked region.  These sources displayed radio source counts consistent with those found for the VLA-COSMOS 1.4 GHz survey \citep{bondi2008}.  Full details of the observations, data reduction, and source catalogue can be found in XXL Paper XI.

\begin{table}
\caption{ATCA configurations and corresponding observing times for XXL-S.}
\centering
\begin{tabular}{c c c}
\small Dates & \small Configuration & \small Net Observing Time (hours)\\
\hline
\hline
3-5 Sep 2012 & 6A & 29.3\\
25 Nov 2012 & 1.5C & 12.9\\
8-12 Nov 2014 & 1.5A & 41.4\\
16-30 Dec 2014 & 6A & 136.2\\
\hline
 & Total & 219.8\\
\end{tabular}
\label{tab:observations}
\end{table}

\subsection{New observations of XXL-S}

The ATCA observations of the remaining 20 deg$^2$ were completed in 2014 using the same frequency coverage as the pilot survey.  Over a total of 240 hours, 390 pointings were observed.  To maximise efficiency, the pointings were arranged in a hexagonal pattern such that the separation between pointing centres was the FWHM of the primary beam at 2.1 GHz ($\sim$12.3$'$) for Nyquist sampling.  Each day's schedule was designed to maximise the number of $uv$ samplings for each pointing, which were scanned for 60 seconds each.  The number of $uv$ samplings for a given pointing ranged from 22-28, with 42\% of the pointings having the average number of 25 $uv$ samplings, and 76\% of the pointings having between 24 and 26 $uv$ samplings.  In order to achieve the highest resolution possible, 75\% of the data was taken in the extended 6A configuration.  The remaining 25\% of the data was taken in the more compact 1.5A configuration to gain additional $uv$ coverage and surface brightness sensitivity.
Thus, each pointing had $\sim$3-5 times as many 6A observations as 1.5A configurations.  This was designed to allow sensitivity to extended structure while still gaining a sufficient resolution to be able to cross-match the sources to other multiwavelength data for source classification.  

The primary flux and bandpass calibrator, PKS 1934-638 \citep{reynolds1994}, was observed for 10 minutes at the beginning of each observing run.  During each run, the secondary phase calibrator, PKS 2333-528, was also observed to determine the antenna complex gains and polarisation leakage correction in the calibration step (see the next section).  This source was scanned for 2 minutes every 20-25 minutes.

\subsection{Calibration and flagging}

The $uv$ data were calibrated by using the Multichannel Image Reconstruction, Image Analysis and Display (\textsc{miriad}) software \citep{sault1995}.  The following process was repeated for each day of observation.  The channels in the CABB that are known to be affected by self-interference from 640 MHz clock harmonics were removed with the ``birdie'' option in the \texttt{ATLOD} task.    The task \texttt{MFCAL} was then used on the primary calibrator (1934-638) in order to determine the gains, delay terms, and passband responses for each ATCA antenna for all the frequencies across the bandwidth of the receivers (2048 MHz).  As part of this process, \texttt{PGFLAG}, which is based on \texttt{AOFLAGGER} \citep{offringa2010,offringa2012}, was used to flag the primary calibrator for radio frequency interference (RFI).  Next, \texttt{GPCAL} was used to determine the antenna gains and phases and the polarisation leakages as a function of polarisation axis for each antenna, in 8 frequency bins (each 256 MHz wide).

This calibration was then applied to the secondary calibrator (2333-528) using \texttt{GPCOPY}.  \texttt{PGFLAG} was used to flag RFI and \texttt{GPCAL} was used to determine the antenna gains, phases and polarisation leakages that applied to the secondary.  \texttt{GPBOOT} was then used to correct the gains by comparing them to that of the primary.  Lastly, \texttt{MFBOOT} was used to correct the secondary's flux density by comparing the observations to its known flux density.

Once these steps were finished, the calibration of the primary and secondary calibrators were copied to the $uv$ data for each pointing using \texttt{GPCOPY}.  To complete the calibration of the data, RFI was flagged in the $uv$ data for each pointing with \texttt{PGFLAG}.

\subsection{Imaging}
\label{sec:imaging}

An image of each pointing was constructed from its corresponding $uv$ data.  Two different imaging techniques were tested: imaging in the full 2.0 GHz-wide band (the ``full-band'') and in eight 256 MHz-wide sub-bands.  The main reason for doing the latter is the significant variations of the primary beam response, the synthesised beam, and the flux density of most sources across the large fractional bandwidth of $\Delta \nu / \nu=(2.0 \text{ GHz} / 2.1 \text{ GHz})\sim1$.  After a number of tests and comparisons, the full-band mosaic was chosen as the final image from which the sources would be extracted (\citealp{novak2015,smolcic2017_submitted}).  The primary reason for this was a systematic downward trend away from 1.0 in the plot showing the integrated to peak flux density ratio $S_{\rm{int}}/S_{\rm{p}}$ vs signal-to-noise ratio $S/N$ (defined as $S_{\rm{p}}/\sigma$, where $\sigma$ is the rms noise) for the sub-band combined mosaic.  The trend would have introduced significant systematic uncertainties into the source flux densities (see Appendix~\ref{sec:sb} for the details).

The procedure for imaging each pointing was similar to that of the pilot survey and is as follows.  The following steps were executed twice for all 491 pointings in order to minimise $\sigma_{\rm{ind}}$ (the rms noise for an individual pointing):

\begin{enumerate}
\item{\texttt{INVERT} the $uv$ data to generate its inverse Fourier transform (the ``dirty'' image).  A robust parameter \citep{briggs1995} of 0.0 was used.}
\item{\texttt{MFCLEAN} \citep{sault1994} down to 10$\sigma_{\rm{ind}}$ to get the ``clean'' model of the dirty image that corresponds to a peak residual of less than 10$\sigma_{\rm{ind}}$.}
\item{\texttt{SELFCAL} the $uv$ data to improve the phase calibration with the model generated by \texttt{MFCLEAN}.}
\end{enumerate}

The final image was then cleaned down to 6$\sigma_{\rm{ind}}$, and \texttt{RESTOR} was used to convolve it with the clean beam (the best fit to the main lobe of the dirty beam).  The last step was to convolve each pointing's final restored image to the smallest elliptical Gaussian possible. This ensured that all the pointings had the exact same resolution across the mosaic.  This final Gaussian beam size was $5.39'' \times 4.21''$, and its position angle was 2.4$^{\circ}$.  Therefore, the geometric mean of the final resolution of the mosaic was $\sqrt{B_{\text{maj}}B_{\text{min}}} = \sqrt{5.39'' \times 4.21''} \approx 4.8''$.

\subsection{Peeling}

Radio telescopes are sensitive to sources in off-axis directions due to the multi-lobe structure of the antenna response patterns (i.e. the ``sidelobes'').  This causes the images of areas on the sky near a bright ($\sim$1 Jy) source to have increased noise due to the response of the sidelobes to the nearby bright source.  There are seven bright sources in the XXL-S radio mosaic that significantly increased the noise in their vicinity, as indicated by the number and morphology of the image artefacts surrounding them.

The noise in the images surrounding the seven bright sources was improved by a process called ``peeling'' (e.g. \citealp{intema2009}).  The following steps were performed for each of those pointings:

\begin{enumerate}
\item Image the pointing according to the procedure in Section~\ref{sec:imaging}.
\item Subtract all sources found in the pointing's 6$\sigma_{\rm{ind}}$ \texttt{MFCLEAN} model 
from the pointing's $uv$ data using the task \texttt{UVMODEL}.
\item Image the bright source by offsetting the pointing's $uv$ data to the direction of the bright source.  Three loops of \texttt{INVERT}, \texttt{MFLCEAN} with 100 iterations, \texttt{RESTOR}, and \texttt{SELFCAL} were used to generate a model of the bright source.  This ensured that the \texttt{SELFCAL} solutions would be better for the direction of the bright source.
\item Subtract the bright source from the $uv$ data with \texttt{UVMODEL}, using the model generated in step 3.
\item Invert the gains acquired from the bright source in step 3 using \texttt{GPEDIT} and apply those gains with \texttt{UVAVER}.
\item Add the sources back into the $uv$ data with \texttt{UVMODEL}, using the model from step 2.
\item Redo the imaging as described in Section~\ref{sec:imaging}.
\end{enumerate}

Peeling effectively removes the gains acquired from the bright sources and replaces them with only the gains from the other sources in the pointing, hence minimising the effect of the sidelobes.  A comparison of the peeled noise histogram and the unpeeled noise histogram (Figure \ref{fig:noise_plot}) demonstrates that this had the effect of reducing the number of pixels containing relatively high rms values ($\sigma>50$ $\mu$Jy) and increasing the number of pixels containing relatively low rms values ($\sigma<50$ $\mu$Jy).  The number of pointings that were peeled was 164.

\subsection{Mosaic image}

Once all the individual images were produced, the mosaic of the full XXL-S field was constructed by using \texttt{LINMOS}.  This task applies the primary beam correction, which is the inverse of the primary beam response as a function of radius, at 10 different frequency ranges across the 2 GHz bandwidth to each image before combining the images.

\begin{figure*}
        \includegraphics[width=\textwidth]{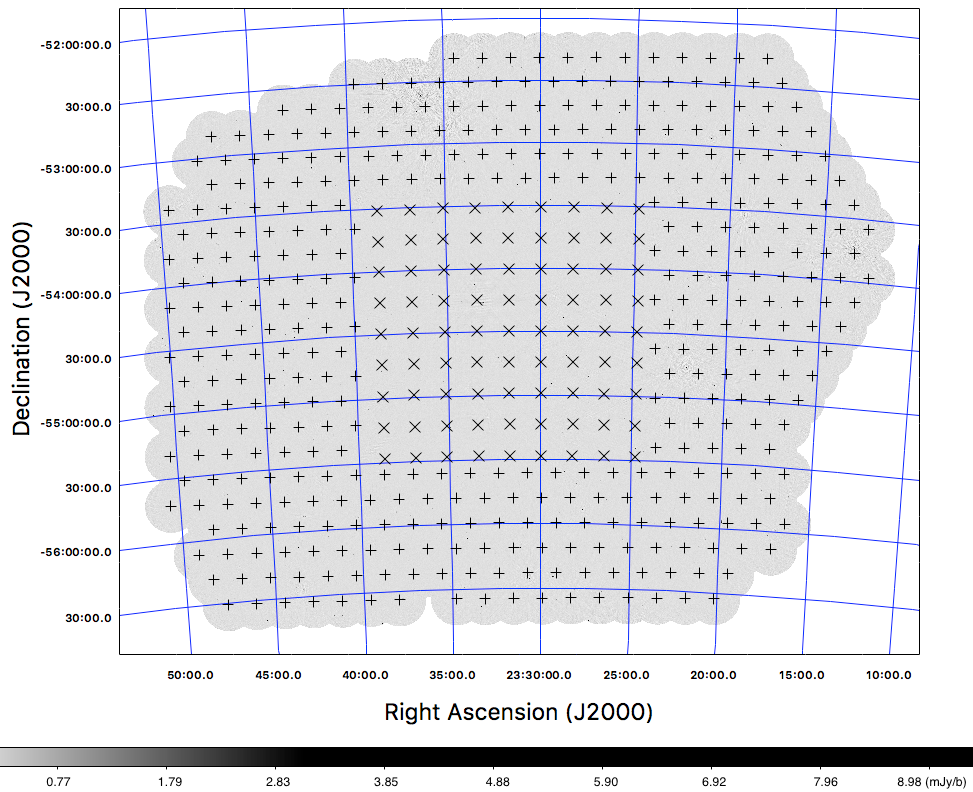}
    \caption{2.1 GHz mosaic of XXL-S, displayed with an inverted greyscale shown at the bottom (with units of mJy/beam) and the RA and Dec coordinate grid overlaid.  The ATCA pointing centres are also overlaid, where the pilot pointings are marked by `$\times$' and the 2014 pointings are marked by `+'.  The edges have been masked to exclude pixels more than 12.3$'$ (the FWHM of the primary beam at 2.1 GHz) away from the pointing centres.}
    \label{fig:mosaic_pointings}
\end{figure*}

\begin{figure}
        \includegraphics[width=\columnwidth]{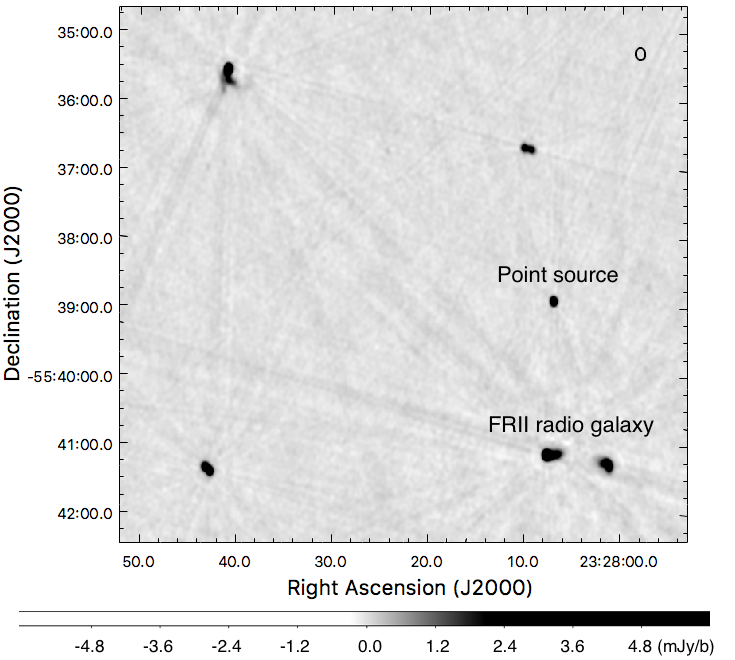}
    \caption{Zoomed-in section of the 2.1 GHz mosaic of XXL-S, displayed with an inverted greyscale shown at the bottom (with units of mJy/beam).  One point source and one definite FRII radio galaxy are labeled.  The beam size is shown at the upper right corner.}
    \label{fig:mosaic_zoom}
\end{figure}

Figure \ref{fig:mosaic_pointings} shows the final XXL-S ATCA 2.1 GHz mosaic overlaid with the ATCA pointing patterns.  The edges have been masked to exclude pixels more than 12.3$'$ (the FWHM of the primary beam at 2.1 GHz) away from the pointing centres.  This was required because the noise becomes non-Gaussian near the edges, and hence it would have significantly contributed to the noise tail (see Section~\ref{sec:noise}), making source finding unreliable in those regions.  The total area of the final masked mosaic is 23.32 deg$^2$.  Figure \ref{fig:mosaic_zoom} shows a zoomed-in portion of the mosaic to demonstrate the typical quality of the image.  The zoom-in contains one point source and a definite Fanaroff and Riley II (FRII) radio galaxy \citep{fanaroff1974}.

\begin{table*}
\caption{Properties of the ATCA XXL-S observations.  The largest angular scale is the largest well-imaged structure the ATCA was sensitive to (at 1.809 GHz, the median effective frequency).  The largest source in XXL-S is $\sim$105$''$ across on the sky.}
\centering
\begin{tabular}{c c c c c c c}
 & & Primary beam FWHM & & & & Largest angular scale\\
Frequency coverage & Channels & (at 2.1 GHz) & Resolution & Median noise & Area & (at 1.809 GHz)\\
\hline
\hline
1.076 $-$ 3.124 GHz & 2048 & 12.3$'$ & 4.76$''$ & 41.3 $\mu$Jy/beam & 23.32 deg$^2$ & 140$''$\\
\end{tabular}
\label{tab:obs_summary}
\end{table*}

\section{Properties of final mosaic}
\label{sec:mosaic_properties}

\subsection{Noise properties}
\label{sec:noise}

In order to extract sources from the mosaic, an rms noise estimate needed to be calculated for each pixel.  The source extractor software \textsc{aegean} \citep{hancock2012} was used to accomplish this task by creating a separate mosaic designated as the noise map.  \textsc{aegean} constructs the noise map by dividing up the mosaic image into areas that have the size of 20x20 synthesised beams ($108'' \times 84''$ or $154 \times 120$ pixels, in this case).  For each of these areas in the mosaic image, it calculates the rms noise based on the interquartile range (IQR) as rms = IQR/1.349\footnote{\scriptsize \url{http://www.physics.usyd.edu.au/~hancock/files/AegeanUserGuide.pdf}}.  \textsc{aegean} then assigns this rms value to every pixel in the corresponding area of the noise map, and it continues until all areas in the mosaic image have been covered.

 \begin{figure}
	\includegraphics[width=\columnwidth]{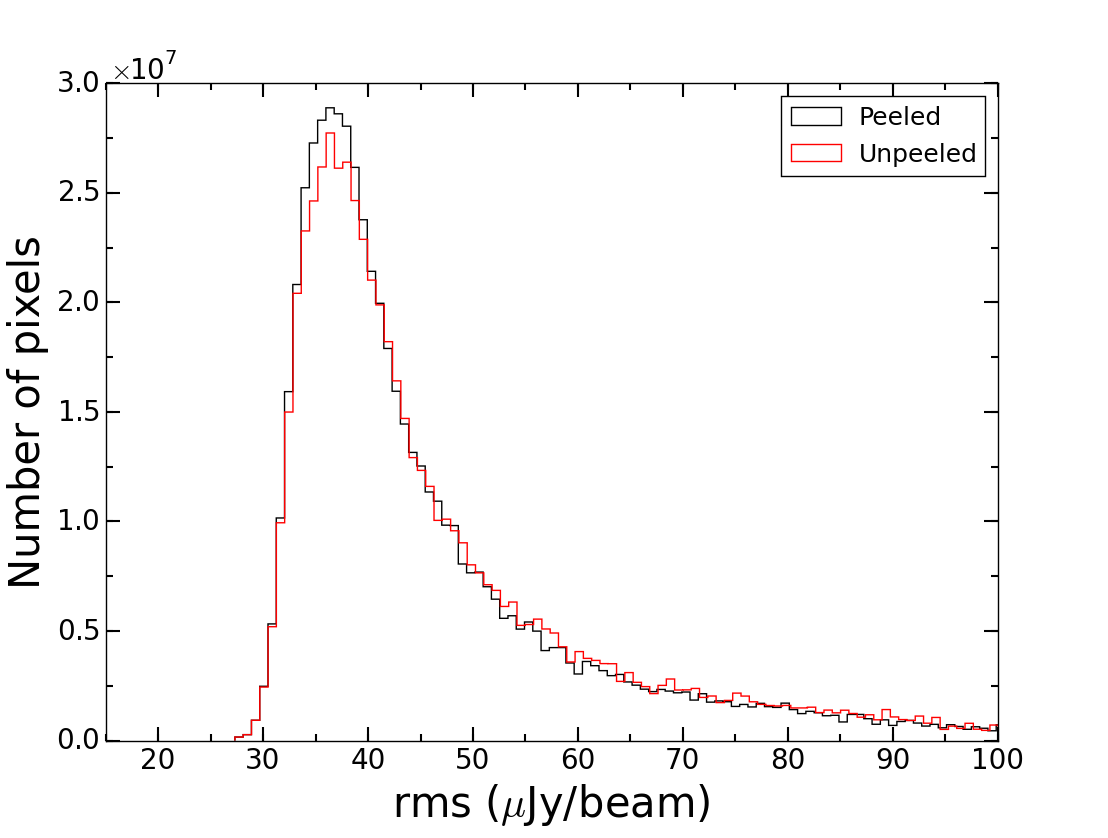}
    \caption{Noise distribution for the final 2.1 GHz mosaic of XXL-S (black histogram).  The red histogram is the rms noise distribution for the same mosaic with the peeling step excluded.  This demonstrates that the peeling reduced the number of high-noise pixels and increased the number of low-noise pixels.}
    \label{fig:noise_plot}
\end{figure}

Figure~\ref{fig:noise_plot} shows the distribution of rms noise values in the final noise map (black histogram), which peaks at 36.5 $\mu$Jy/beam.  The median occurs at 41.3 $\mu$Jy/beam, with a minimum and maximum of 27.3 $\mu$Jy/beam and 658.0 $\mu$Jy/beam, respectively.  Given the large 25 deg$^2$ area observed in XXL-S, the median rms sensitivity of $\sigma \approx41$ $\mu$Jy/beam means that this is the largest area radio survey conducted down to these flux density levels.  Table \ref{tab:obs_summary} shows a summary of the properties of the observations.

\subsection{Source extraction}

\textsc{blobcat} \citep{hales2012} was used to extract the radio sources in XXL-S.  \textsc{blobcat} is based on the ``flood-fill'' algorithm and was specifically designed to perform efficient source extraction on large radio survey data.  A detection threshold of 5$\sigma$ and a ``floodclip'' threshold of 3$\sigma$ were used.  Due to the complex morphology of radio sources, the following definitions are adopted:

\begin{enumerate}
\item{Blob -- a group of pixels in the mosaic image whose fitted peak, as determined by \textsc{blobcat}, is above the detection threshold and extends to all surrounding pixels that are above the ``floodclip'' threshold.}
\item{Component -- the radio emission that gives rise to a blob.}
\item{Source -- one or more components that are associated with the same galaxy.}
\end{enumerate}

\begin{figure}
	\includegraphics[width=\columnwidth]{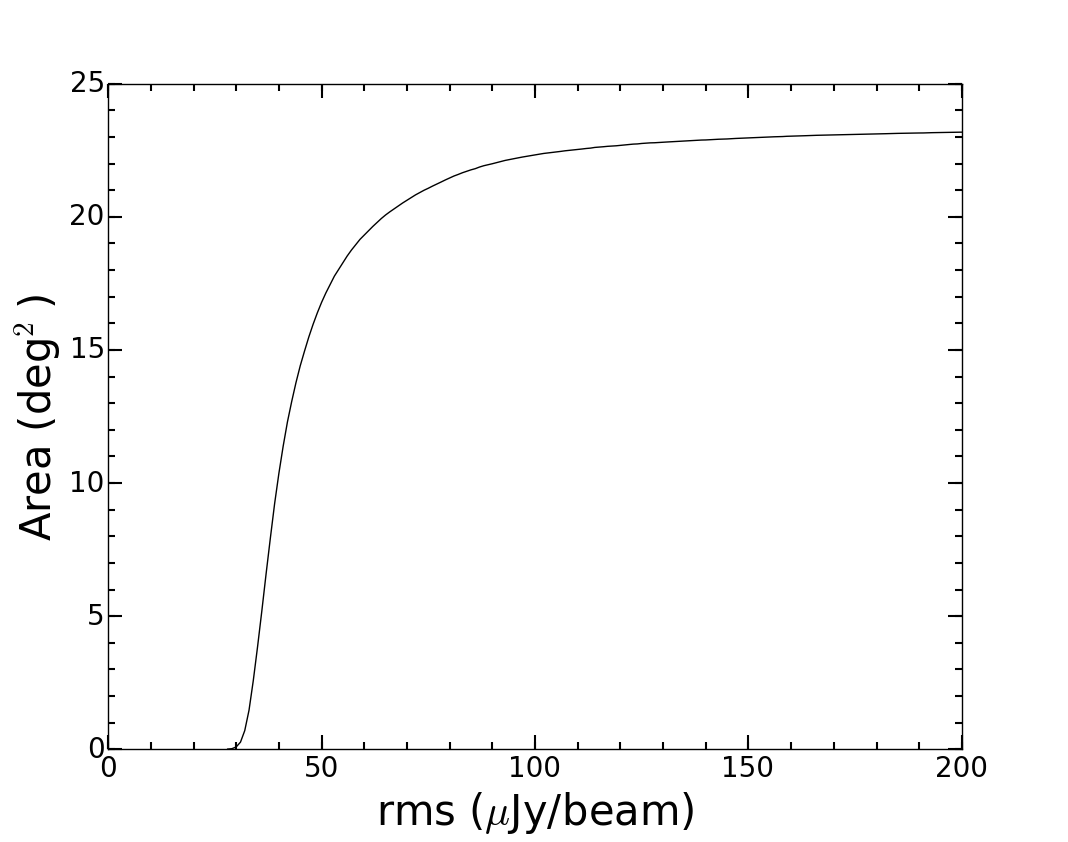}
    \caption{Visibility area (mosaic sensitivity function) plot for the final 2.1 GHz mosaic of XXL-S, indicating the total area with a noise level below a given rms.}
    \label{fig:vis_area_plot}
\end{figure}

In other words, ``blobs'' represent what the software detects, but ``components'' and ``sources'' represent real galaxies that the software is attempting to describe.  Running \textsc{blobcat} with the above settings resulted in an extraction of 6482 blobs.  Figure \ref{fig:vis_area_plot} shows the visibility area as a function of $\sigma$, which indicates the total area in the mosaic with a noise level below a given rms.

\subsection{Spurious blobs}

Some of the blobs were clearly image artefacts caused by antenna sidelobes.  A blob was considered spurious if it was very clear that it was due to the sidelobes in the vicinity of a source brighter than $S_{\rm{p}}>30$ mJy (e.g. a ``clean stripe'').  These blobs were visually searched for in the mosaic, and 141 ($\sim$2.2\%) were found and removed from the initial catalogue.

\begin{figure}
	\includegraphics[width=\columnwidth]{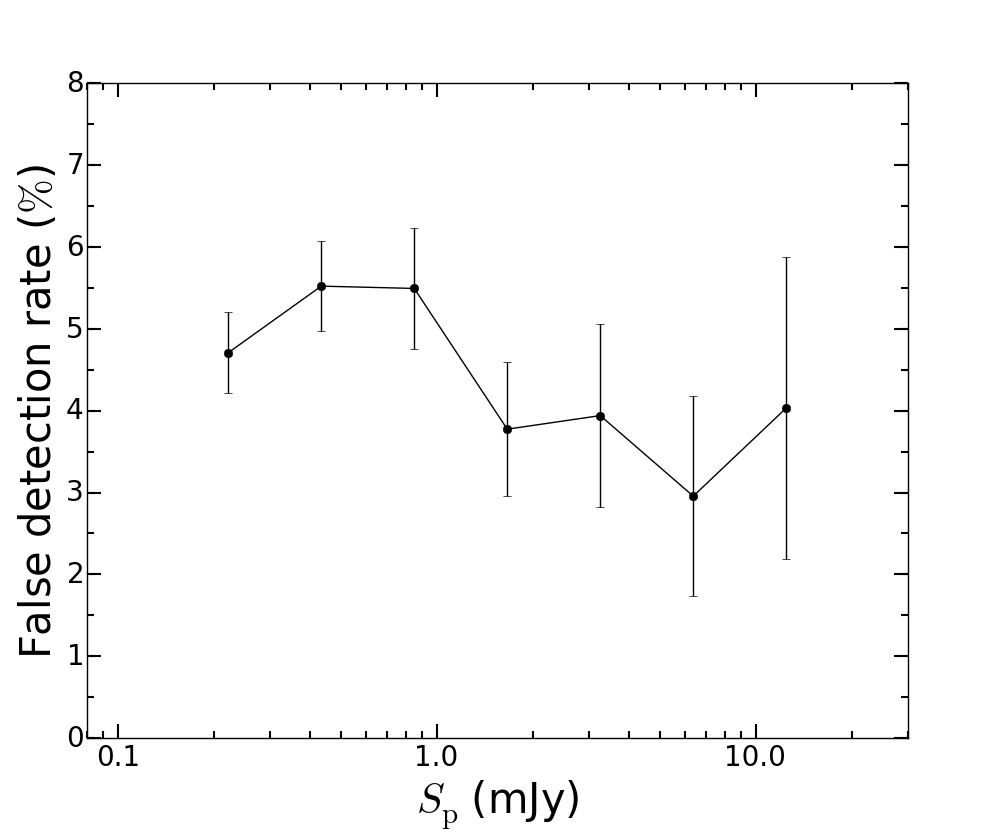}
    \caption{False detection percentage vs $S_{\rm{p}}$ bin.  For each of the 7 $S_{\rm{p}}$ bins, the false detection percentage is defined as $(N_{\rm{F}}/N_{\rm{T}})\times100$, where $N_{\rm{F}}$ is the number of blobs that were detected in the negative mosaic and $N_{\rm{T}}$ is the number of blobs that were detected in the final mosaic.  The uncertainties are defined according to Equation \ref{eq:fdr_uncert}.  There were no false detections for $S_{\rm{p}}>17$ mJy.}
    \label{fig:false_detections_plot}
\end{figure}

The empirical false detection rate was also determined.  Since the rms noise distribution contains both positive and negative peaks, the negative peaks can be used to estimate the false detection rate.  A negative mosaic was created by multiplying all the pixel values in the mosaic image by $-1$.  When \textsc{blobcat} was run on this negative mosaic, it detected 304 blobs.  All false detections were characterised by $S_{\rm{p}}<17$ mJy, and the percentage of false detections as a function of peak flux density bin $S_{\rm{p}}$ is shown in Figure \ref{fig:false_detections_plot}.  For each $S_{\rm{p}}$ bin, the false detection rate percentage (FDR\%) is defined as $(N_{\rm{F}}/N_{\rm{T}})\times100$, where $N_{\rm{F}}$ is the number of blobs that were detected in the negative mosaic and $N_{\rm{T}}$ is the number of blobs that were detected in the final mosaic.  A Poissonian uncertainty in the false detection rate percentage  ($\delta$FDR\%) was calculated according to:
\begin{equation}
\label{eq:fdr_uncert}
\delta $FDR\% = (FDR\%)$ \sqrt{\left(\sqrt{N_{\rm{F}}}/N_{\rm{F}}\right)^2 + \left(\sqrt{N_{\rm{T}}}/N_{\rm{T}}\right)^2}.
\end{equation}
At the faintest $S_{\rm{p}}$ bin ($S_{\rm{p}} < 0.3$ mJy), the false detection percentage is $\sim$4.75\%.  For the $S_{\rm{p}}$ bins brighter than this, the percentage fluctuates between $\sim$3.0\% and $\sim$5.5\%.

Spread over the whole catalogue, the empirical false detection rate was 4.7\%.  If the 141 actual spurious blobs that were found by a manual search are subtracted from the 304 false detections, 163 spurious blobs are expected to be missed.  Therefore, the maximum overall false detection rate is approximately 2.5\%.  After the 141 manually identified spurious blobs were removed, the total number of blobs (and components) was 6341.

\subsection{Deconvolution}
\label{sec:deconvolution}

The measurement of the angular extent of a radio source is directly defined by the ratio of its integrated flux $S_{\rm{int}}$ to its peak flux $S_{\rm{p}}$:
\begin{equation}
\frac{S_{\rm{int}}}{S_{\rm{p}}} = \frac{\theta_{\rm{maj}} \theta_{\rm{min}}}{B_{\rm{maj}} B_{\rm{min}}}
\end{equation}
Here, $\theta_{\rm{maj}}$ and $\theta_{\rm{min}}$ are the major and minor FWHM of the Gaussian fit to the source and $B_{\rm{maj}}$ and $B_{\rm{min}}$ are the synthesised beam's major and minor FWHM, respectively.  Therefore, this ratio was used to determine whether or not sources are resolved.

\begin{figure}
	\includegraphics[width=\columnwidth]{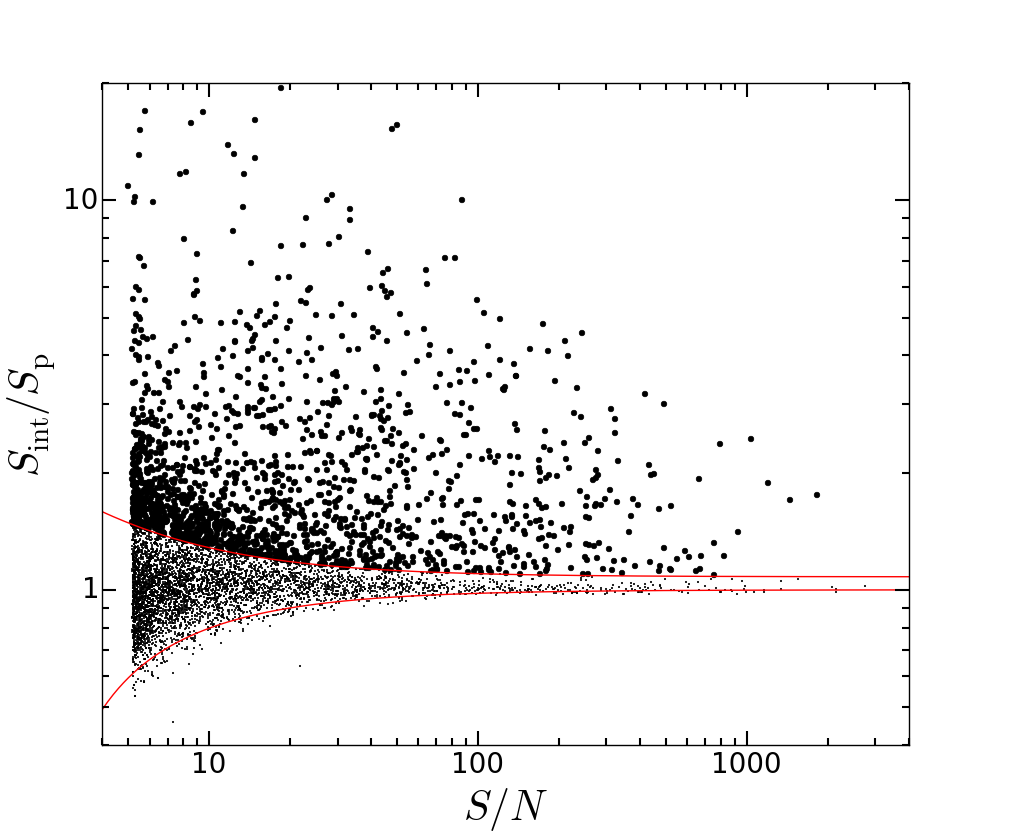}
    \caption{$S_{\rm{int}}/S_{\rm{p}}$ vs. $S/N$ plot for all components with the peak flux densities corrected for bandwidth smearing (see Section~\ref{sec:bws}).   The lower and upper envelopes (red curves) are described by the equations \ref{eq:deconvolution_lower} and \ref{eq:deconvolution_upper}, respectively.  The components above the upper envelope are considered resolved, and the ones below it are considered unresolved.}
    \label{fig:Sint_Sp_SNR_log_plot}
\end{figure}

Figure \ref{fig:Sint_Sp_SNR_log_plot} shows the $S_{\rm{int}}/S_{\rm{p}}$ vs $S/N$ plot for all components.  Instances where $S_{\rm{int}}/S_{\rm{p}}<1$ are due to the component size uncertainty introduced by the noise in the image.  These uncertainties must be taken into account when determining whether or not a component is resolved.  Following, for example, \cite{prandoni2000}, \cite{bondi2003}, \cite{huynh2005}, \cite{bondi2008}, and \cite{huynh2015}, the empirical function describing the lower edge of the data points in the $S_{\rm{int}}/S_{\rm{p}}$ vs $S/N$ plot was defined such that 90\% of the components with $S_{\rm{int}}/S_{\rm{p}}<1$ were above it.  This curve is described by the equation:
\begin{equation}
\label{eq:deconvolution_lower}
S_{\rm{int}} / S_{\rm{p}} = 1.00 - \frac{2.03}{S/N}.
\end{equation}
This function was then mirrored above the x-axis.  However, Figure \ref{fig:Sint_Sp_SNR_log_plot} shows a number of components with $S/N>100$ very close to $S_{\rm{int}} / S_{\rm{p}}=1$.  Most of these components are within $\sim$5\% of the beam size, which is within the calibration error.  There are also a handful of $S/N$>1000 components near $S_{\rm{int}} / S_{\rm{p}}=1$ that, upon visual inspection, appeared point-like.  Therefore, in order for all these components to be considered unresolved, and to reduce the slight peak bias introduced by the manual bandwidth-smearing correction (see Section \ref{sec:bws}), the offset was increased to 1.08 for the upper envelope.  This criteria is similar to the requirement in \cite{franzen2015} that $S_{\rm{int}} / S_{\rm{p}}$ be greater than 1.15 for a high $S/N$ source to be considered resolved.  The equation for the upper curve is therefore:
\begin{equation}
\label{eq:deconvolution_upper}
S_{\rm{int}} / S_{\rm{p}} = 1.08 + \frac{2.03}{S/N}.
\end{equation}
Both of these curves are shown as the red lines in Figure~\ref{fig:Sint_Sp_SNR_log_plot}.  Components that are above the upper red curve are considered resolved, and components below it are considered unresolved.  In the component catalogue (Section \ref{sec:final_cats}), the ``resolved'' flag is set to 0 for unresolved components and 1 for resolved components, and the deconvolved sizes of unresolved components are assigned a value of zero.  The number of resolved components is 1677 out of 6350, or 26.4\% of the total.  This is broadly consistent with the expected value based on the percentage of resolved components in other radio surveys: VLA-COSMOS found 28.6\% with 1.5$''$ resolution \citep{bondi2008}, the pilot survey found 22\% with $\sim$4.4$''$ resolution (XXL Paper XI), ELAIS-S1 found 19\% with $\sim$9.8$''$ resolution \citep{franzen2015}, and CDFS found 17\% with $\sim$10.6$''$ resolution \citep{franzen2015}.

\subsection{Bandwidth smearing}
\label{sec:bws}

The initial $S_{\rm{int}}/S_{\rm{p}}$ vs. $S/N$ plot that was generated indicated that the median value of $S_{\rm{int}}/S_{\rm{p}}$ for components with $S/N>50$ was 1.035.  If the median $S_{\rm{int}}/S_{\rm{p}}$ is offset above 1.0, it indicates that point sources appear larger than the beam, which is normally caused by bandwidth smearing (e.g. \citealp{bondi2008}).  Bandwidth smearing occurs when the visibilities of a source measured in a finite channel bandwidth $d \nu$ are treated as though they correspond to the central frequency $\nu_0$ of the channel \citep{bridle1999}.  This frequency averaging results in the image of the source being smeared in a radial direction away from the phase centre, decreasing $S_{\rm{p}}$ while keeping $S_{\rm{int}}$ constant \citep{condon1998}.  Therefore, the peak flux density of every component was corrected for bandwidth smearing by multiplying them all by 1.035.  Figure \ref{fig:Sint_Sp_SNR_log_plot} has this correction in place.

\subsection{Complex blobs}
\label{sec:complex}

\subsubsection{Gaussian fitting with \texttt{IMFIT}}
\label{sec:imfit}

\begin{figure}
	\includegraphics[width=\columnwidth]{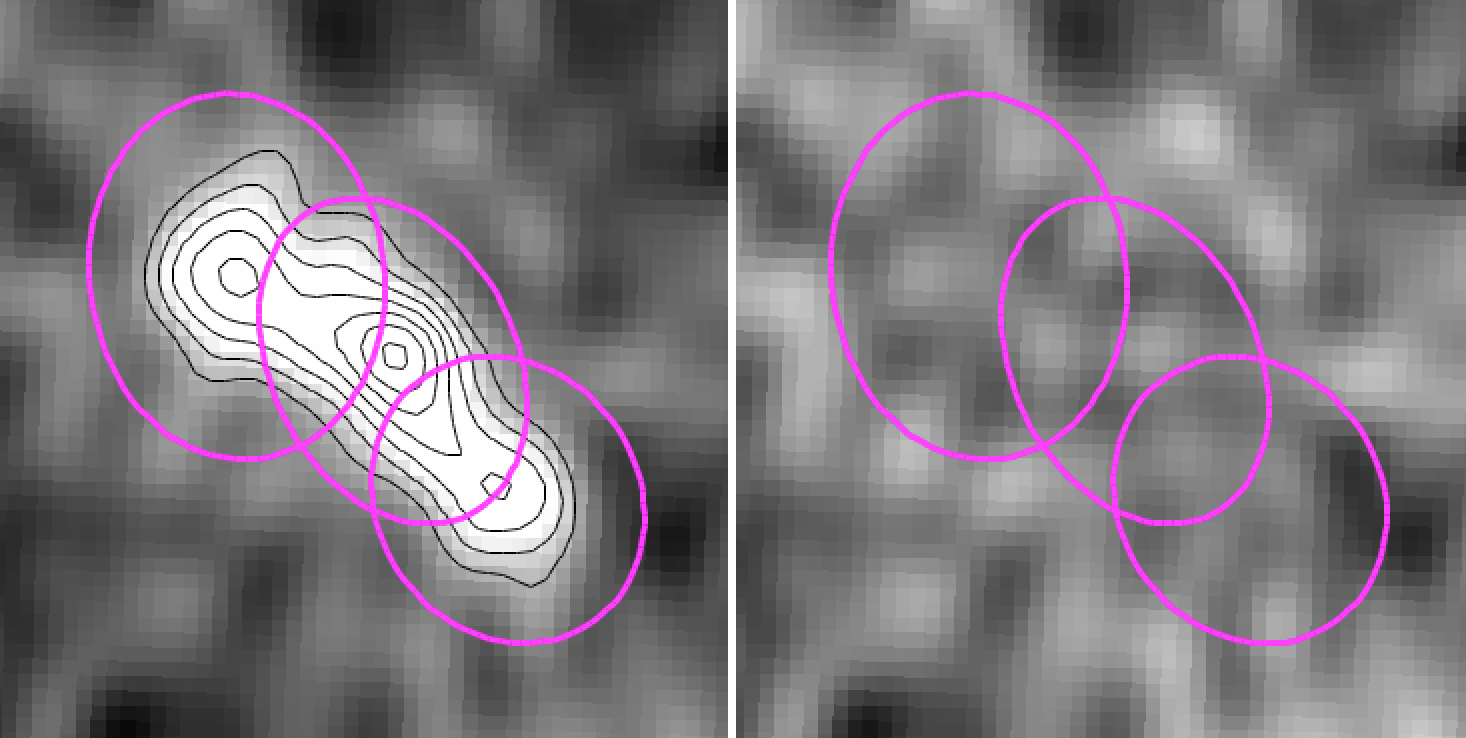}
    \caption{Example of a complex blob.  The left panel shows the blob itself with the 2.1 GHz 3-20$\sigma$ contours (in steps of $\sim$2.4$\sigma$) in black and the 2D elliptical Gaussian fits in magenta.  The right panel shows the same ellipses overlaid on the residual image.  The length of the major and minor axes of the middle elliptical Gaussian fit are 8.07$''$ and 5.78$''$, respectively.}
    \label{fig:complex_source}
\end{figure}

Some components extracted by \textsc{blobcat} exhibited complex non-Gaussian morphology and therefore needed to have their flux densities and positions confirmed.  Following the recommendation of \cite{hales2012} and the criteria of \cite{franzen2015}, components were defined as complex if they had $N_{\rm{pix}}>300$ and $R^{\rm{EST}}>1.4$.  $N_{\rm{pix}}$ is the number of pixels included in the blob and $R^{\rm{EST}}$ is the ratio between the measured sky area of the
detected blob and the sky area of an unresolved Gaussian blob with the same $S_{\rm{p}}$.  During this stage of analysis, a floodclip threshold ($T_f$) of 2.6$\sigma$ was used to extract the flux densities of the complex blobs to determine if this was appropriate for the XXL-S data.  The number of complex components from the $T_f=2.6\sigma$ catalogue was 610, or $\sim$10\% of the total.  The \textsc{miriad} task \texttt{IMFIT} was used to fit them.  For each component, an elliptical Gaussian was added until the peak in the residual image was less than 5$\sigma$, following \cite{franzen2015}.  Figure \ref{fig:complex_source} shows an example of a complex component and the fit that was necessary to achieve a peak residual below 5$\sigma$.  

After the fitting was completed, it was found that 139 of these complex components required only one elliptical Gaussian to achieve an acceptable fit.  Therefore, as recommended by \cite{hales2012}, the data from the original \textsc{blobcat} entries for these 139 components were kept in the $T_f=2.6\sigma$ catalogue.  For the remaining 471 blobs, their elliptical Gaussian fits found with \texttt{IMFIT} were combined into one catalogue entry.  This involved summing the flux densities of the elliptical Gaussian fits (using $S_{\rm{p}}$ if the Gaussian was unresolved and $S_{\rm{int}}$ if it was resolved) and calculating their flux-weighted positions.

\subsubsection{Comparison of \texttt{IMFIT} Gaussian fits and \texttt{BLOBCAT} measurements}

\begin{figure}
	\includegraphics[width=\columnwidth]{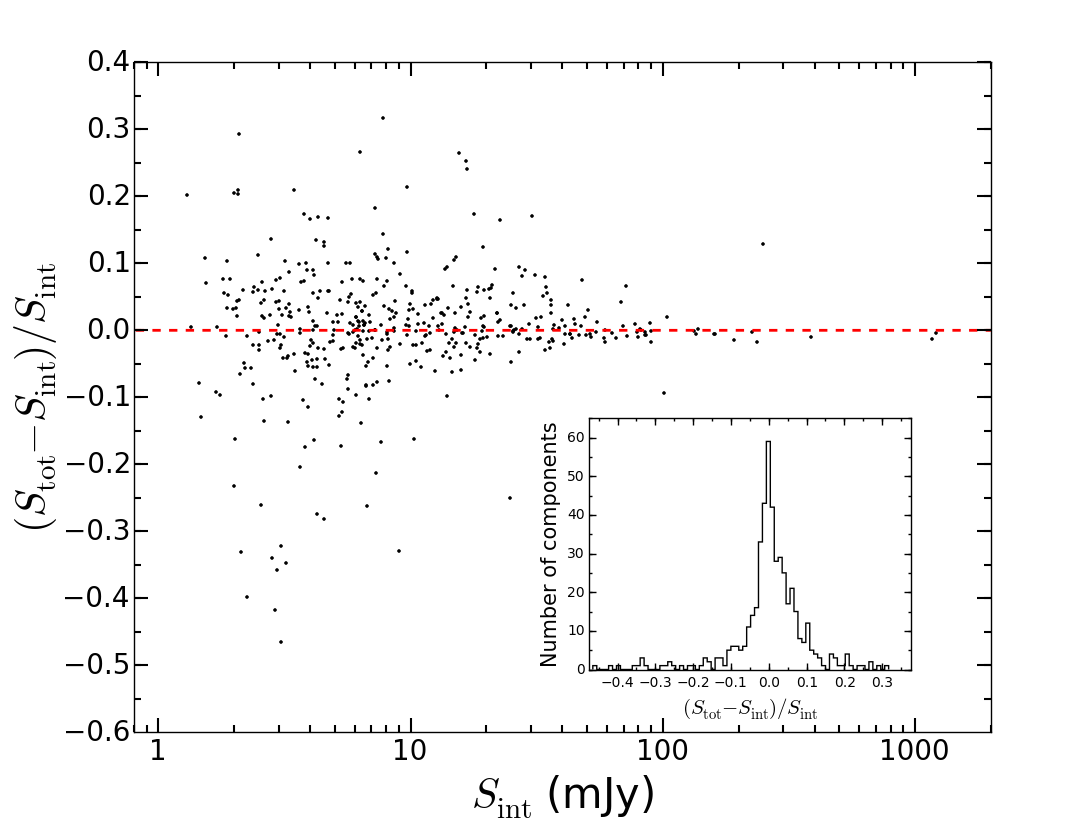}
    \caption{Fractional difference between $S_{\rm{tot}}$ from IMFIT and $S_{\rm{int}}$ from \textsc{blobcat} as a function of $S_{\rm{int}}$ for complex components in the $T_f=2.6\sigma$ catalogue.  The dashed red line shows where $(S_{\rm{tot}}-S_{\rm{int}})/S_{\rm{int}}=0$ to help guide the eye. The inset plot shows the histogram of the $(S_{\rm{tot}}-S_{\rm{int}})/S_{\rm{int}}$ values, the median and standard deviation of which are 0.005 and 0.092, respectively.}
        \label{fig:Stot_Sint_diff_ratio_plot}
\end{figure}

Figure \ref{fig:Stot_Sint_diff_ratio_plot} shows the fractional difference between the \texttt{IMFIT} total flux densities of the complex components ($S_{\rm{tot}}$) and their integrated flux densities as determined by \textsc{blobcat} ($S_{\rm{int}}$), as a function of $S_{\rm{int}}$.  The median value was 0.005 and the standard deviation was 0.095, meaning that the complex components had fractional differences within 10\%.  However, there were clearly some outliers, particularly at $S_{\rm{int}}<10$ mJy.  Visual inspection of these outliers revealed that they featured extended emission with complex morphology that was difficult to capture with Gaussian fitting.  Trying to accurately capture the flux density from complex components with elliptical Gaussian fits is clearly less accurate and reliable than \textsc{blobcat}'s approach of summing the pixels and dividing the sum by the beam size in pixels. Even some brighter components ($S_{\rm{int}}>10$ mJy) were difficult to fit and achieve residuals below 5$\sigma$ because of their complex morphology.  \textsc{blobcat} can handle them much more easily because its floodfill algorithm covers all the pixels.

\begin{figure}
	\includegraphics[width=\columnwidth]{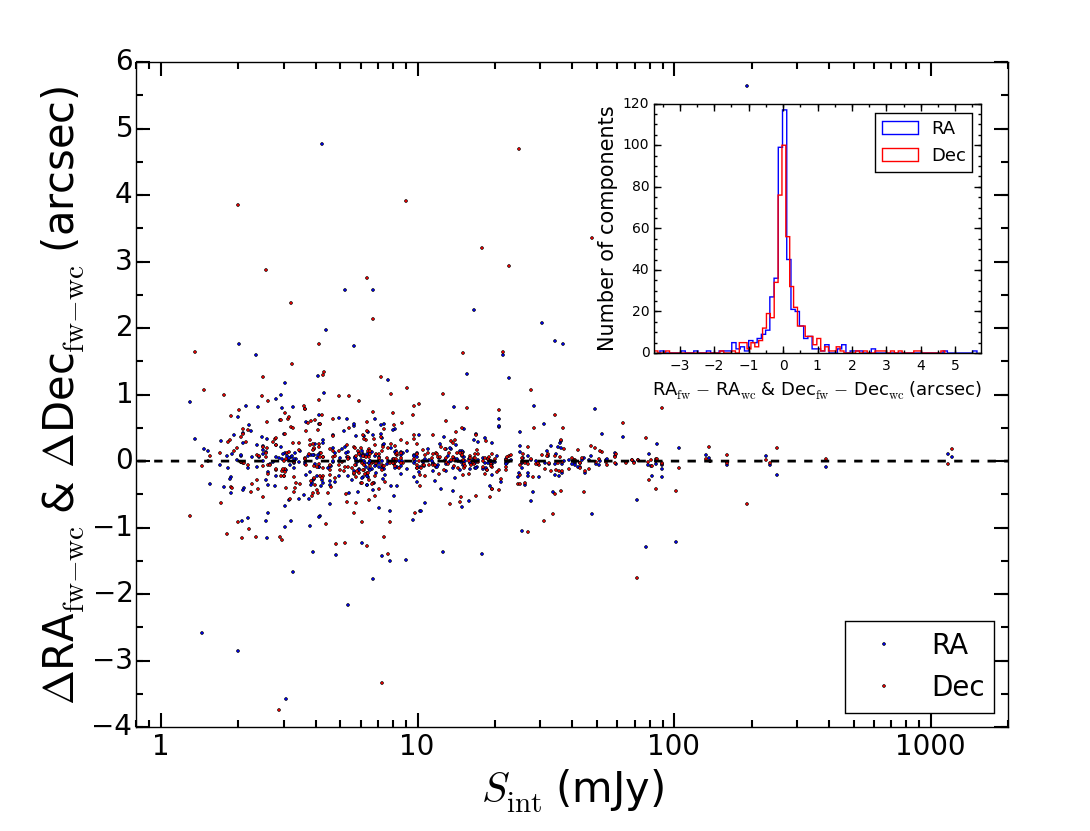}
    \caption{Offsets between \texttt{IMFIT}'s RA$_{\rm{fw}}$ and Dec$_{\rm{fw}}$ and \textsc{blobcat}'s RA$_{\rm{wc}}$ and Dec$_{\rm{wc}}$.  The dashed black line shows where RA$_{\rm{fw}}$ $-$ RA$_{\rm{wc}}=0$ and Dec$_{\rm{fw}}$ $-$ Dec$_{\rm{wc}}=0$ to help guide the eye. The inset plot shows the histogram of the RA$_{\rm{fw}}$ $-$ RA$_{\rm{wc}}$ and Dec$_{\rm{fw}}$ $-$ Dec$_{\rm{wc}}$ values.  The median RA and Dec offsets were $-0.002$ and 0.022 arcseconds and their corresponding standard deviations were 0.668 and 0.689 arcseconds, respectively.}
        \label{fig:RA_Dec_off_Sint_plot}
\end{figure}

\begin{figure*}
	\includegraphics[width=\textwidth]{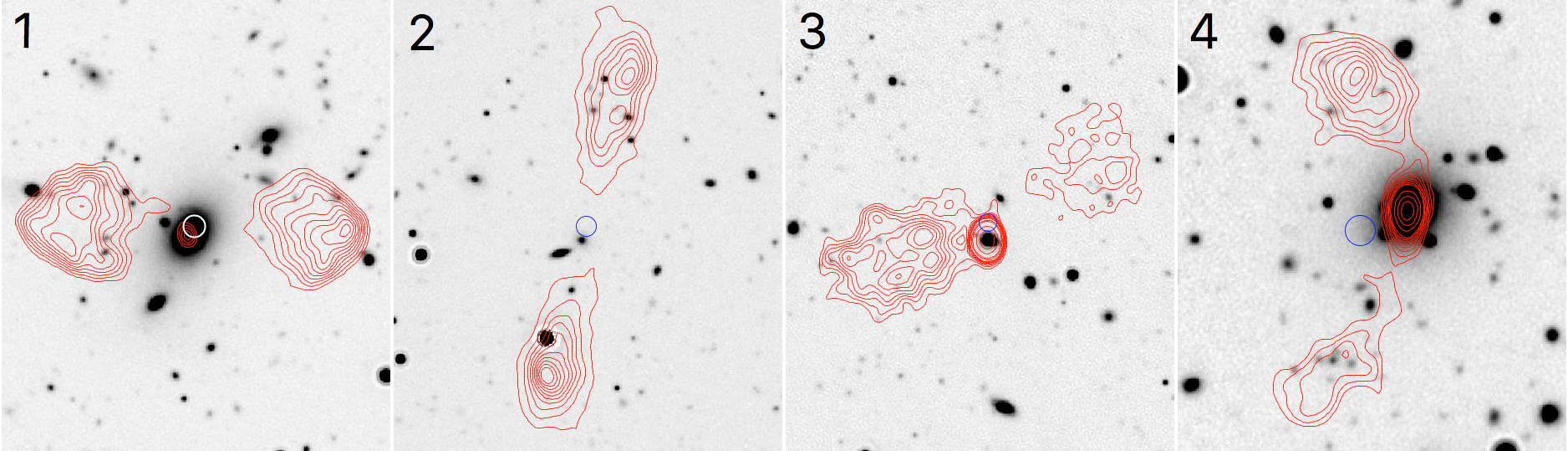}
    \caption{Examples of radio components that were joined into one source in the source catalogue.  The background images are the DECam $z$-band cut-outs, the red contours are the 2.1 GHz contours, and the white/blue circles are the flux-weighted positions of the components with a radius of 2.5$''$.  Panels 1-4 show examples of cases 1-4 as described in Section \ref{sec:gen_source_cat}.  These examples make it clear that the raw flux-weighted positions are not exactly coincident with the obvious $z$-band counterparts. This demonstrates the need to manually adjust the position of these kinds of radio sources (see Section \ref{sec:de-blended}).  The position adjustments will be published in a forthcoming paper of this series (Butler et al., in prep).}
    \label{fig:joined_complex}
\end{figure*}

Figure \ref{fig:RA_Dec_off_Sint_plot} shows the positional offsets ($\Delta$RA$_{\rm{fw-wc}}$ and $\Delta$Dec$_{\rm{fw-wc}}$) between the flux-weighted RA and Dec of the \texttt{IMFIT} Gaussian fits (RA$_{\rm{fw}}$ and Dec$_{\rm{fw}}$) for the complex components
and \textsc{blobcat}'s flux-weighted position measurements (RA$_{\rm{wc}}$ and Dec$_{\rm{wc}}$) of the complex components.  The median RA and Dec offsets were $-0.002$ and 0.022 arcseconds, and their corresponding standard deviations were 0.668 and 0.689 arcseconds, respectively.  Therefore, the majority of components had RA and Dec offsets within $\pm$0.7 arcsec.  The few extreme outliers were either, again, faint components with extended emission that were difficult to model with Gaussian fits, or bright sources with particularly complex morphology.  For example, the isolated data point in the upper right of the plot in Figure \ref{fig:RA_Dec_off_Sint_plot} is due to an obvious local FRII radio galaxy.

The \textsc{blobcat} measurements of $S_{\rm{int}}$ and the flux-weighted positions RA$_{\rm{wc}}$ \& Dec$_{\rm{wc}}$ were solely used because the \textsc{blobcat} flux density and position measurements of the complex components were similar to that of the majority of the \texttt{IMFIT} Gaussian fits and some of the \texttt{IMFIT} Gaussian fits were clearly inadequate.  In addition, the floodclip threshold of 2.6$\sigma$ sometimes captured more noise than real flux density for pixels close to this threshold, especially for the fainter complex components.  Therefore, to maximise the capturing of actual flux density from the sources and to minimise the addition of noise, a floodclip threshold of 3$\sigma$ was used for the final catalogue.  The number of complex blobs in the final component catalogue was 501 (7.9\% of the total).

\subsection{De-blending complex blobs into separate components}
\label{sec:de-blended}

The positions of radio components associated with optical galaxies are frequently offset from the optical source positions.  As a result, the flux-weighted positions \textsc{blobcat} generates for complex blobs are not always suitable for finding their optical counterparts, which is a major step in the science goals of this series of papers.  To obtain accurate positions of the complex blobs, their radio contours were overlaid onto DECam \citep{flaugher2015} $z$-band images and examined by eye.  The $z$-band images had a depth of $z<24.5$ and were processed with an improved pipeline used to process the Blanco Cosmology Survey dataset \citep{desai2012}.  A technique for single-component sources analogous to the one described in Section \ref{sec:gen_source_cat} was utilised to determine the most likely $z$-band counterpart, which in most cases was the one closest to the raw flux-weighted position.  The position of each complex blob was adjusted accordingly.  The component and source catalogues presented in Section \ref{sec:final_cats} and Tables \ref{tab:comp_catalogue} and \ref{tab:source_catalogue} contain only the raw flux-weighted positions.  The position adjustments of the complex sources will be published in a forthcoming paper of this series (Butler et al., in prep).

In addition, separate radio sources along the same line of sight are sometimes blended together in the same blob, which gives rise to complex morphology.  For a given complex blob, if two optical sources clearly appeared to be coincident with exactly one radio peak each or two optical sources were the clear counterparts to two different regions of radio emission within the blob, that blob was de-blended into two separate components.  This process was performed for seven blobs using the procedure described in Section \ref{sec:imfit}.  Each of these blobs was de-blended into two or three elliptical Gaussians.  For five out of the seven blobs, two elliptical Gaussians corresponded to one component each.  For the other two blobs, three elliptical Gaussians were required.  In both of these cases, a single Gaussian corresponded to one component and two Gaussians corresponded to the other component. Since the latter were not sufficiently modelled with a single Gaussian fit, they are considered complex and resolved components.  In order to calculate each of their $S_{\rm{int}}$ values, the $S_{\rm{int}}$ of the single Gaussian component from their corresponding blob was subtracted from \textsc{blobcat}'s $S_{\rm{int}}$ value for the entire blob.  The positions of the optical counterparts of these two complex components were chosen as their radio positions.

In the component catalogue (Section \ref{sec:final_cats}), the IDs of these components start with the blob number as originally catalogued by \textsc{blobcat} followed by ``-\#'', where \# is a number assigned to that component (e.g. blob 337 was de-blended into the components labelled as 337-1 and 337-2).  This process increased the total number of components to 6348.

\subsection{Combining components into radio sources}
\label{sec:gen_source_cat}

In some cases, \textsc{blobcat} catalogued two separate components, but in fact they form part of the same radio source such as an FRII radio galaxy.  In order to find these sources, each complex component was cross-matched to all other complex components within 150$''$.  Furthermore, all non-complex components were cross-matched to two complex components within 150$''$ (to cover FRI and FRII radio galaxies with cores), and any additional non-complex components were cross-matched to complex components within 45$''$.  Each of these groups of matches were examined by eye.

DECam $z$-band imaging was used in the same way as in Section \ref{sec:de-blended} to aid in the determination of whether or not separate radio components belonged to the same radio galaxy.  The following four basic cases were considered to be one radio source:

\begin{enumerate}
\item Three radio components (e.g. two lobes and a core) were aligned with each other (i.e. elongated along the same axis) and there was clearly one optical source near the flux-weighted position (within $\sim$2-5$''$).  An example is shown in panel 1 of Figure \ref{fig:joined_complex}.  If there were optical counterparts for the core and only one of the other components, the three components were still considered one source.  If all three had an optical counterpart, and the morphology was not indicative of a typical radio galaxy, then they were all considered separate radio sources.
\item Two aligned radio components had no optical counterparts, but there was an optical source in between the two near the flux-weighted position.  Panel 2 of Figure \ref{fig:joined_complex} shows one example of this case.  If there was an optical counterpart for one of the components but not the other, the two components were still considered one source (unless the component with a counterpart appeared to be an unresolved source).  If both had an optical counterpart, and there was no indication of the typical morphology of radio galaxies, then they were considered separate radio sources.
\item Two radio components, one with emission that appeared point-source-like in one area and extended in another area, were aligned with each other.  The component with extended emission had an optical counterpart at the peak of its flux density distribution.  In this case, the core of the radio emission was blended with one of the lobes, resulting in just one component near the core.  In Figure \ref{fig:joined_complex}, such an example is shown in panel 3.  If both components had an optical counterpart, then they were considered separate sources.
\item Two or more components were configured on the sky in such a way that they were linked by a curved path, and there was an optical counterpart at the ``head'' of the curve.  This case covered radio galaxies with bent radio jets, most of which are called wide angle tail (WAT) radio galaxies, which are usually shaped by ram pressure as the galaxy travels through the intracluster medium.  Panel 4 in Figure \ref{fig:joined_complex} shows an example of a WAT radio galaxy.  This is also an example of case 3 except for the fact that the components are not aligned.  If there was no optical counterpart near the head of the WAT in these cases, then the components were considered separate radio sources.
\end{enumerate}

If a given group of components satisfied one of the criteria above, it was considered a ``multiple-component'' source.  The catalogue entries of their components were combined into one source by summing their flux densities and calculating their flux-weighted positions.  Additionally, their peak flux densities, $S/N$, and deconvolved sizes were set to -99.  This process was repeated for all multiple-component sources (see Section \ref{sec:final_cats}).

There were four special cases of multiple-component sources that clearly featured the morphology of a typical radio galaxy and had a clear $z$-band counterpart, but significant portions of their flux densities (up to 50\%) were missing because \textsc{blobcat} did not detect one of the blobs involved due to the fact that the missed blob had $S/N<5$.  For these four sources, \textsc{blobcat} was re-run using a detection threshold of 4$\sigma$, which detected the missing blobs.  They were then added to the component catalogue and included in the calculations for their corresponding sources as described in the previous paragraph.  The radio galaxy shown in panel 3 of Figure \ref{fig:joined_complex} is one of the four sources for which this was done (the right blob was added to the source).

Interestingly, there is a three-component giant radio galaxy at the very north edge of the mosaic.  Unfortunately, the masked mosaic cut off part of its western component, which had ~$S_{\rm{p}}<4\sigma$.  Given the latter and the fact that giant radio galaxies are not involved in the main science goals of this series of papers, the two blobs found for this source were deleted from the catalogue.

\subsection{Final number of radio components and sources}

The number of components in the final component catalogue is 6350.  Out of these, 111 were combined into 48 multiple-component sources, leaving 6239 single-component sources.  The final source catalogue thus contains a total of 6287 sources, out of which 1626 (25.9\%) were resolved.

\subsection{Completeness simulation}
\label{sec:completeness_sim}

\begin{figure}
	\includegraphics[width=\columnwidth]{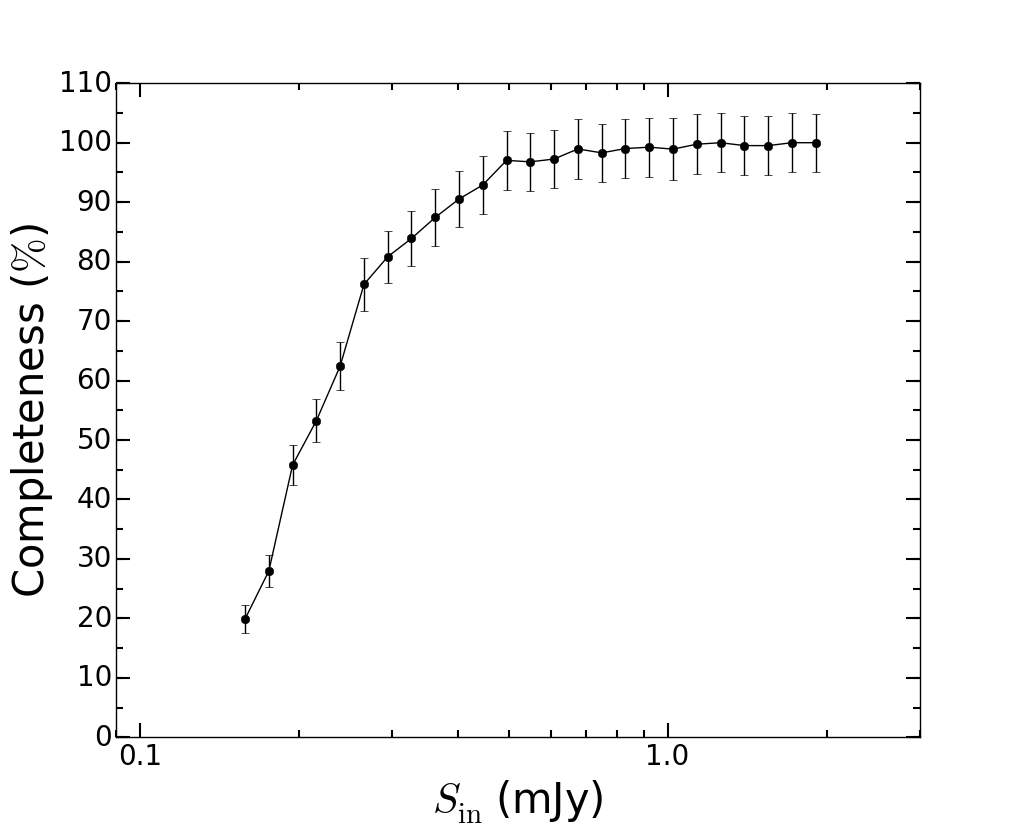}
    \caption{Completeness curve for the XXL-S field, showing completeness percentage ($100 \times N_{\rm{det}}/N_{\rm{tot}}$) vs. input flux density ($S_{\rm{in}}$).  The percentage uncertainties for each $S_{\rm{in}}$ bin are Poissonian (i.e. $100 \times \sqrt{N_{\rm{det}}}/N_{\rm{tot}}$).}
    \label{fig:completeness_plot}
\end{figure}

It is important to understand the detection rate of sources as a function of flux density.  The use of a simple detection threshold above the local noise leads to incompleteness at faint flux density levels, which must be taken into account for deriving radio source counts and radio luminosity functions.  This incompleteness can be characterised by a Monte Carlo simulation.  To perform the simulation, 10,000 point sources of random flux density from 0.15 to 2 mJy were injected, one at a time, into random positions across the mosaic using the \textsc{miriad} task \texttt{IMGEN}.  \textsc{blobcat} was then executed on those positions with the same input parameters that were used to detect the real radio sources in the mosaic.  The percentage of sources that were detected by \textsc{blobcat} ($100 \times N_{\rm{det}}/N_{\rm{tot}}$) was calculated for 25 flux density bins from 0.15 to 2 mJy.  This is designated as the ``completeness percentage'' ($CP$).  Figure \ref{fig:completeness_plot} shows the plot of $CP$ vs. input flux density bin ($S_{\rm{in}}$), with the percentage Poissonian uncertainties of $100 \times \sqrt{N_{\rm{det}}}/N_{\rm{tot}}$.  At about 0.9 mJy, essentially 100\% completeness is reached, and 95\% completeness is achieved at about 475 $\mu$Jy.  At the nominal 5$\sigma$ detection limit of $\sim$200 $\mu$Jy, the completeness is $\sim$45\%.  These numbers are taken into account in the calculation of the source counts (Section \ref{sec:source_counts}).

\begin{figure}
	\includegraphics[width=\columnwidth]{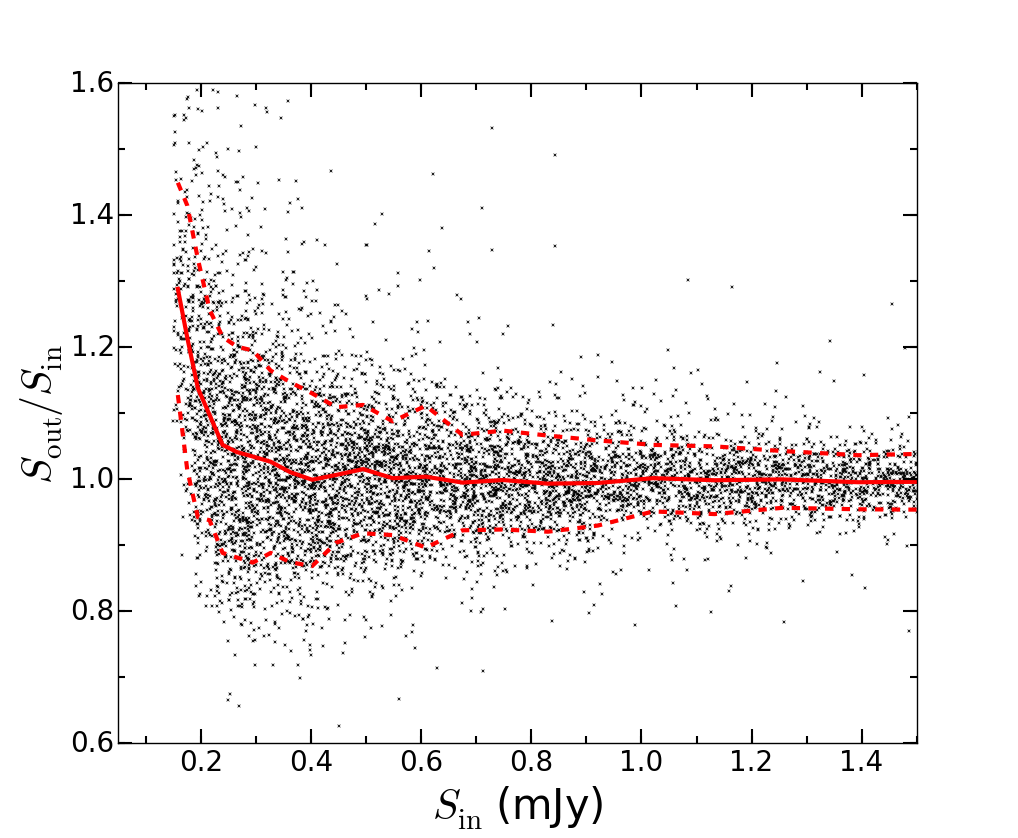}
    \caption{Plot of ratio of output flux density to input flux density ($S_{\rm{out}}/S_{\rm{in}}$) vs input flux density ($S_{\rm{in}}$) in the completeness simulation.  The $S_{\rm{in}}$ bins are the same as in Figure~\ref{fig:completeness_plot}.  The solid red line represents the median $S_{\rm{out}}/S_{\rm{in}}$ of the $S_{\rm{in}}$ bins, and the dashed red lines represent the $\pm$1$\sigma$ $S_{\rm{out}}/S_{\rm{in}}$ bounds.}
    \label{fig:s_frac_plot}
\end{figure}

\begin{figure*}
	\includegraphics[width=\textwidth]{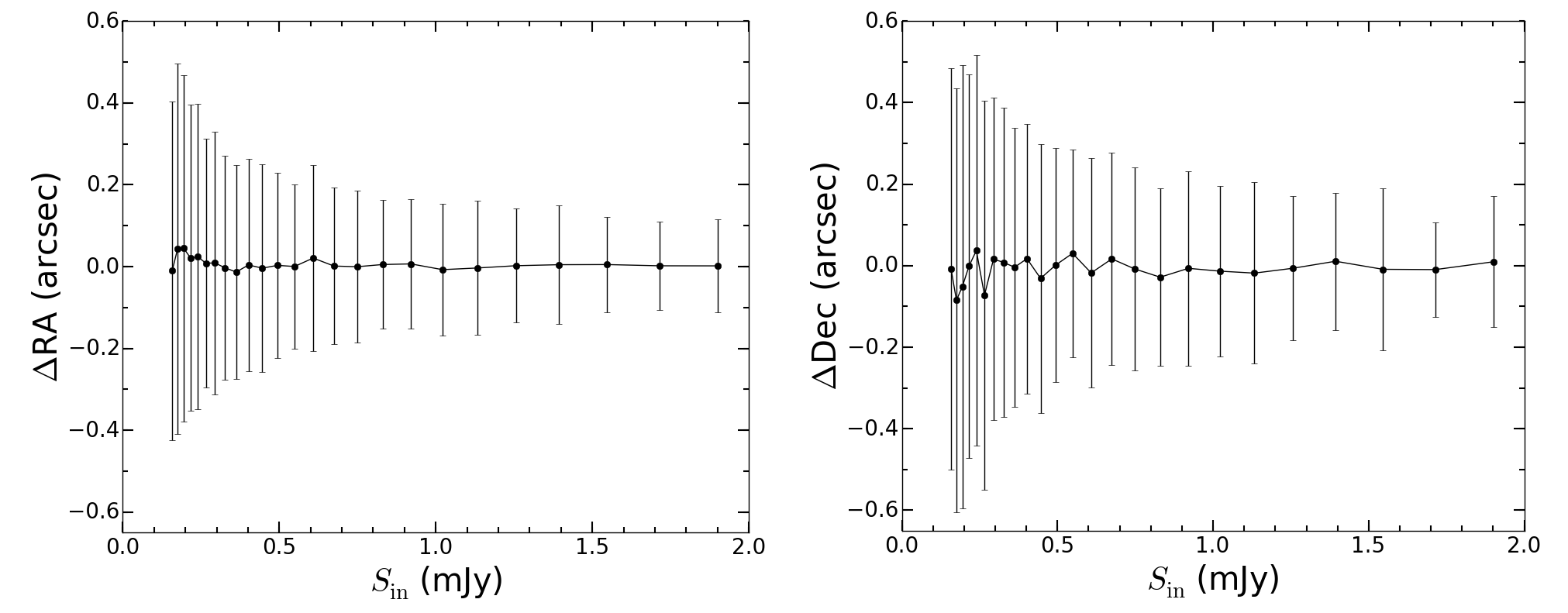}
    \caption{Plots of RA offset (left panel) and Dec offset (right panel) between the output positions as determined by \textsc{blobcat} and the input positions ($\Delta$RA and $\Delta$Dec, respectively) as a function of input flux density ($S_{\rm{in}}$) in the completeness simulation.  The black dots are the median $\Delta$RA and $\Delta$Dec for the same bins as in Figure \ref{fig:completeness_plot}, and the error bars are the $\pm$1$\sigma$ $\Delta$RA and $\Delta$Dec bounds.}
    \label{fig:del_RA_del_Dec_plot}
\end{figure*}

\subsubsection{Flux density boosting}
\label{subsec:flux_boosting}

Sources that happen to lie on a noise peak have larger flux densities than they otherwise would, and therefore they have a higher chance of being detected.  On the other hand, sources that are located on a noise trough have smaller flux densities and a decreased probability of being detected.  This effect can lead to an artificial flux density boosting, which has the largest effect on the faintest detectable sources.  The magnitude of flux density boosting can be estimated from the ratio of output flux density to input flux density, $S_{\rm{out}}/S_{\rm{in}}$.  This is illustrated in Figure~\ref{fig:s_frac_plot}, which shows the plot of $S_{\rm{out}}/S_{\rm{in}}$ vs $S_{\rm{in}}$.  The flux densities are boosted by about 30\% for $\sim$150 $\mu$Jy, and by about 5\% for $\sim$240 $\mu$Jy.  The flux density boosting is negligible (less than 5\%) for flux densities greater than that.  These numbers are also used in the calculation of the radio source counts in Section \ref{sec:source_counts}.

\subsubsection{Positional accuracy}

The completeness simulation also indicates the accuracy to which the positions of the radio sources are recovered by \textsc{blobcat}.  The offset between the input and output RA and Dec positions ($\Delta$RA and $\Delta$Dec, respectively) as a function of $S_{\rm{in}}$ is shown in Figure \ref{fig:del_RA_del_Dec_plot}.  The standard deviation of the RA and Dec offsets in each $S_{\rm{in}}$ bin can be used to calculate the positional accuracy.  At the faintest flux densities ($\sim$150 $\mu$Jy), the RA and Dec uncertainties are 0.4 and 0.5 arcsec, respectively.  For sources with flux densities higher than $\sim$1 mJy, the RA uncertainty is less than 0.17 arcsec and the Dec uncertainty is less than 0.21 arcsec.

\subsection{\texttt{CLEAN} bias simulation}

\textsc{clean} bias \citep{condon1998} occurs when the \texttt{CLEAN} algorithm reassigns flux density from low $S/N$ sources to noise peaks in the image, which results in a systematic underestimation of the flux density of real sources.  This occurs because some noise peaks may be slightly above the \textsc{clean} threshold.  As a result, when the \textsc{clean} algorithm finds these peaks, it assigns flux density to those peaks and takes away flux density from \textsc{clean} components of real sources.  The underlying cause of this bias is an incomplete $uv$ coverage, and its magnitude is a function of $uv$ coverage and the flux density level down to which the images are cleaned.

A \textsc{clean} bias simulation was performed on a pointing with an average $uv$ coverage.  One point source of random flux density between 0.215 mJy and 1.075 mJy (4$\sigma$ and 20$\sigma$ for the pointing, respectively) was injected into the $uv$ data at a random position within 20 arcmin of the phase centre.  Since \texttt{SELFCAL} had already been performed, the final imaging steps of \texttt{INVERT}, \texttt{MFCLEAN} down to 6$\sigma$, \texttt{RESTOR}, and \texttt{CONVOL} to a common beam size of $5.39'' \times 4.21''$ were then executed.  This process was repeated 10,000 times. 

The output flux density of each simulated source was measured by evaluating the interpolated image pixel value at the injected position.  This was done because the thermal noise fluctuations cause a slight shift in the peak position of a source.  At low $S/N$, this effect is more pronounced because the noise level is a higher percentage of the peak flux density.  Since source extractors like \textsc{blobcat} measure the peak flux density to detect sources, the flux density measured will always be higher than the true flux density of a given low $S/N$ source.  Therefore, measuring the peak flux density of a source will result in flux densities that are slightly positively biased.  On the other hand, the flux density at the injected position of a source will be unbiased on average \citep{franzen2015}.  Therefore, to separate this peak flux density bias and \textsc{clean} bias, the output flux densities were measured at their injected positions, not the peak positions of the output sources.

\begin{figure}
	\includegraphics[width=\columnwidth]{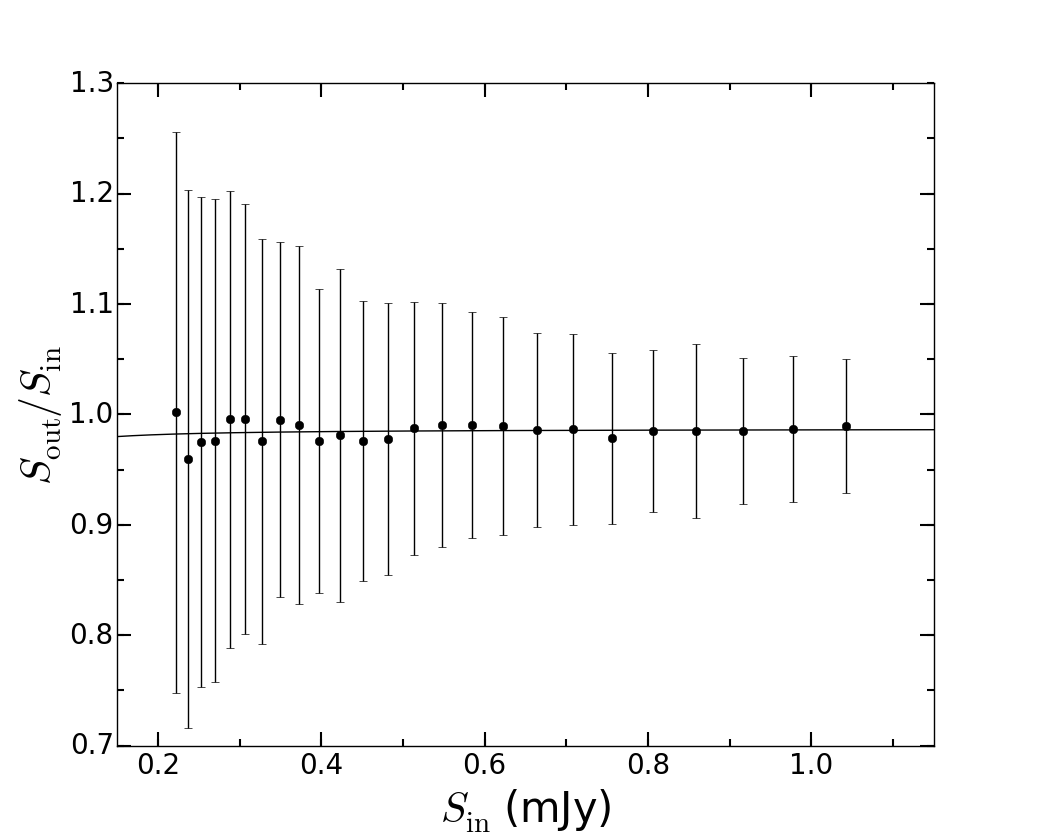}
    \caption{Plot of the ratio of measured output flux density to input flux density ($S_{\rm{out}}/S_{\rm{in}}$) as a function of input flux density ($S_{\rm{in}}$) in the \textsc{clean} bias simulation.  The black dots are the median $S_{\rm{out}}/S_{\rm{in}}$ for each $S_{\rm{in}}$ bin and the error bars represent $\pm$1$\sigma$ in $S_{\rm{out}}/S_{\rm{in}}$ for each bin.  The black line represents the best fit function of the data points, which is $S_{\rm{out}}/S_{\rm{in}} = 0.987 - 19.927/(S_{\rm{in}}/\sigma_{\rm{rms}})$, where $\sigma_{\rm{rms}}$ is the local rms noise. The plot indicates that \textsc{clean} bias is negligible.}
    \label{fig:clean_bias_plot}
\end{figure}

The ratio of output to input flux density ($S_{\rm{out}}/S_{\rm{in}}$) for all 10,000 sources was calculated.  For each of 25 input flux density bins ($S_{\rm{in}}$), the median $S_{\rm{out}}/S_{\rm{in}}$ and the standard deviation of $S_{\rm{out}}/S_{\rm{in}}$, representing the error bars, was determined.  Figure \ref{fig:clean_bias_plot} shows a plot of the median $S_{\rm{out}}/S_{\rm{in}}$ vs $S_{\rm{in}}$ and the best fit function to the data points of 
\begin{equation}
S_{\rm{out}}/S_{\rm{in}} = 0.987 - \frac{19.927}{S_{\rm{in}}/\sigma_{\rm{rms}}},
\end{equation}
where $\sigma_{\rm{rms}}$ is the local rms noise.  It indicates that the \textsc{clean} algorithm underestimates the flux density of a $\sim$0.215 mJy (4$\sigma$) source by no more than 5\%.  This is within the calibration error of 5\%, indicating that \textsc{clean} bias is negligible.  Therefore, the source flux densities were not corrected for \textsc{clean} bias.

\section{Radio component and source catalogues}
\label{sec:final_cats}

Two 2.1 GHz radio catalogues were generated for XXL-S:

\begin{enumerate}
\item Component catalogue: catalogue of all 6350 radio components with fitted $S_{\rm{p}}>5\sigma$, including the components from the seven de-blended blobs (Section \ref{sec:de-blended}).
\item Source catalogue: Catalogue of all 6239 single-component sources (including the components from the seven de-blended blobs) plus all 48 multiple-component sources, as described in Section \ref{sec:gen_source_cat}.
\end{enumerate}

Table \ref{tab:comp_catalogue} shows representative entries from the component catalogue, including de-blended components.  Table \ref{tab:source_catalogue} shows representative entries from the source catalogue.  The full catalogues are available as queryable database tables XXL\_ATCA\_16A and XXL\_ATCA\_16B via the XXL Master Catalogue browser\footnote{\url{http://cosmosdb.iasf-milano.inaf.it/XXL}}. Copies will also be deposited at the Centre de Donn\'{e}es astronomiques de Strasbourg (CDS)\footnote{\url{http://cdsweb.u-strasbg.fr}}.  These catalogues supersede the one presented in XXL Paper XI.

The single-component sources in the source catalogue are equivalent to the corresponding entries in the component catalogue.  Therefore, these entries can be considered either components or sources, and for the sake of simplicity, the following description refers only to sources as found in the source catalogue.  Both catalogues have the same column headings, which are described below.

Column 1 lists the identification (ID) number of each source.

Columns 2 and 3 list the flux-weighted RA and Dec, respectively, of each source in degrees.

Columns 4 and 5 contain the RA and Dec uncertainties as calculated by \textsc{blobcat} ($\sigma_{\rm{RA}}$ and $\sigma_{\rm{Dec}}$), in arcseconds.  For an explanation of how these are calculated, see Section 2.6 in \cite{hales2012}.

Column 6 is the local rms noise $\sigma_{\rm{rms}}$ in mJy/beam.

Columns 7 and 8 list the peak flux density ($S_{\rm{p}}$) corrected for bandwidth smearing and the uncertainty in $S_{\rm{p}}$ ($\sigma_{S_{\rm{p}}}$), respectively, in mJy/beam.  The uncertainty $\sigma_{S_{\rm{p}}}$ is defined as the quadrature sum of the absolute calibration error, the pixellation uncertainty, and the local rms noise.  The calibration error was assumed to be 0.05$S_{\rm{p}}$ and the pixellation uncertainty was determined to be 0.01$S_{\rm{p}}$.  See Appendices A and B from \cite{hales2012} for more details.  Multiple-component sources have $S_{\rm{p}}=-99$.

Columns 9 and 10 display the integrated flux density ($S_{\rm{int}}$) and its uncertainty ($\sigma_{S_{\rm{int}}}$), respectively, in mJy.  The uncertainty $\sigma_{S_{\rm{int}}}$ is defined as the quadrature sum of the absolute calibration error and the local rms noise.

Column 11 contains the $S/N$ of each source.  Multiple-component sources have $S/N=-99$.

Column 12 is the resolved flag (abbreviated by ``Res'').  If a source is resolved, this value is 1, and if it is unresolved, this value is 0.  All multiple-component sources are resolved.

Column 13 is the multiple component flag (abbreviated by ``MC'').  If a source has multiple components (or a component is part of a multiple-component source), this value is 1.  Otherwise, this value is 0.

Columns 14 and 15 display the deconvolved source size $\Theta$ and its associated uncertainty $\sigma_{\Theta}$, both in arcseconds.  Unresolved sources have their deconvolved sizes set to 0 arcsec.  Following equations 10 and 11 from \cite{franzen2015}, $\Theta$ and $\sigma_{\Theta}$ are defined as:
\begin{equation}
\Theta=\sqrt{\left(\frac{S_{\rm{int}}}{S_{\rm{p}}} - 1\right) B_{\rm{maj}} B_{\rm{min}}}
\end{equation}
\begin{equation}
\sigma_{\Theta} = \frac{S_{\rm{int}}}{S_{\rm{p}}} \sqrt{ \frac{B_{\rm{maj}}B_{\rm{min}}}{4(S_{\rm{p}} / S_{\rm{int}} - 1)} \left[\left(\frac{\sigma_{S_{\rm{p}}}}{S_{\rm{p}}}\right)^2 + \left(\frac{\sigma_{S_{\rm{int}}}}{S_{\rm{int}}}\right)^2\right]}
\end{equation}

Column 16 shows the effective frequency $\nu_{\rm{eff}}$ of each source at its flux-weighted position in GHz.  For a given radius away from the pointing centre, the effective frequency is defined as the weighted average of the primary beam response across the 2 GHz bandwidth.  Due to the fact that each pointing was imaged out to 8\% of the primary beam response at 2.1 GHz ($\sim$23$'$), the effective frequency will change for each source as a function of distance from the pointing centre.  It is especially true at the edges of the mosaic, where fewer pointings were stacked together.  The effective frequency at which each source's flux density is considered to be valid was determined by constructing a frequency mosaic using \texttt{LINMOS}, setting options=frequency.  For the position of each source in the image mosaic, the effective frequency was read at the corresponding position in the frequency mosaic.  The median effective frequency was $\nu_{\rm{eff}}=1.809$ GHz and its standard deviation was just $\sigma_{\nu_{\rm{eff}}}=0.032$ GHz.

Column 17 is the complex source flag.  If a source is complex according to the definition in Section \ref{sec:imfit}, then its complex flag is set to 1, and 0 otherwise.  All multiple-component sources are complex.  Two of the 14 de-blended sources, 605-1 and 1393-2, are complex.

\section{Spectral indices}
\label{sec:spectral_indices}

To investigate the spectral index ($\alpha$) properties of the XXL-S sources, the Sydney University Molonglo Sky Survey (SUMSS) 843 MHz survey of the entire southern sky \citep{bock1999} was cross-matched to the XXL-S 2.1 GHz source catalogue.  The faintest peak flux density in the SUMSS catalogue \citep{mauch2003} is 6 mJy, so this analysis represents only the brightest sources in XXL-S.  The resolution of SUMSS is $43'' \times 43''\csc|\delta|$, so at the declination of XXL-S ($\delta=-54.5^{\circ}$), the beam size is $43'' \times 52.8''$.  Therefore, each SUMSS source contained in the XXL-S mosaic was cross-matched to XXL-S sources within 52.8$''$.

Since SUMSS has a much lower resolution than XXL-S, the ATCA may have resolved out some of the flux density that was included in the measurements for the SUMSS sources.  In addition, some resolved ATCA XXL-S sources have multiple components with core-jet structures, which tend to have different spectral indices for different regions (e.g. \citealp{dennett-thorpe1999}).  Therefore, in order to simplify the interpretation of the spectral indices, two different cases were examined. The number of unique SUMSS sources matched to exactly one ATCA XXL-S point source (case 1) was 560.  The total number of unique SUMSS sources matched to at least one ATCA XXL-S source, whether those ATCA sources are resolved or unresolved (case 2), is 699, with 872 total matches.  For the latter case, if more than one ATCA XXL-S source was matched to a SUMSS source, the flux densities of the ATCA XXL-S sources were summed to produce the 2.1 GHz flux density of each SUMSS source.

\begin{figure}
	\includegraphics[width=\columnwidth]{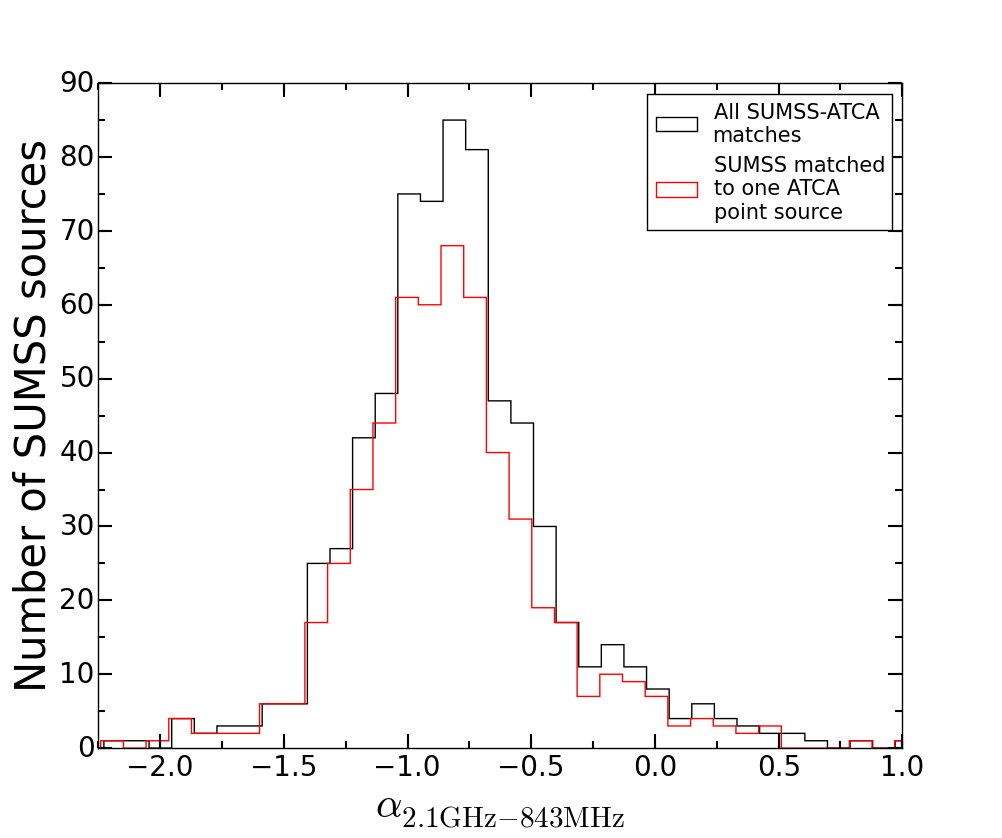}
    \caption{Spectral index distributions for the SUMSS 843 MHz sources cross-matched to one or more XXL-S 2.1 GHz sources.  The black line represents the 699 SUMSS sources cross-matched to all ATCA XXL-S sources within 52.8$''$ and the red line represents the 560 SUMSS sources that were matched to exactly one ATCA XXL-S point source.  The median spectral indices for these cases are $-0.83$ and $-0.85$, respectively.}
    \label{fig:spectral_indices}
\end{figure}

The following convention for radio spectral index $\alpha$ is adopted: $S \propto \nu^{\alpha}$.  Figure \ref{fig:spectral_indices} shows the distribution of spectral indices of the SUMSS sources in the XXL-S field for both cases described in the previous paragraph.  The median spectral index for case 1 (the black line in the figure) is $\alpha=-0.83$ and the standard deviation is $\sigma_{\alpha}=0.50$.  For case 2 (the red line in the figure), the median spectral index is $\alpha=-0.85$ and the standard deviation is $\sigma_{\alpha}=0.48$.  The distributions are centred on very similar median values and have similar shapes, so this indicates that there is little extended emission that the ATCA resolved out for the brightest sources in XXL-S.

\begin{figure}
	\includegraphics[width=\columnwidth]{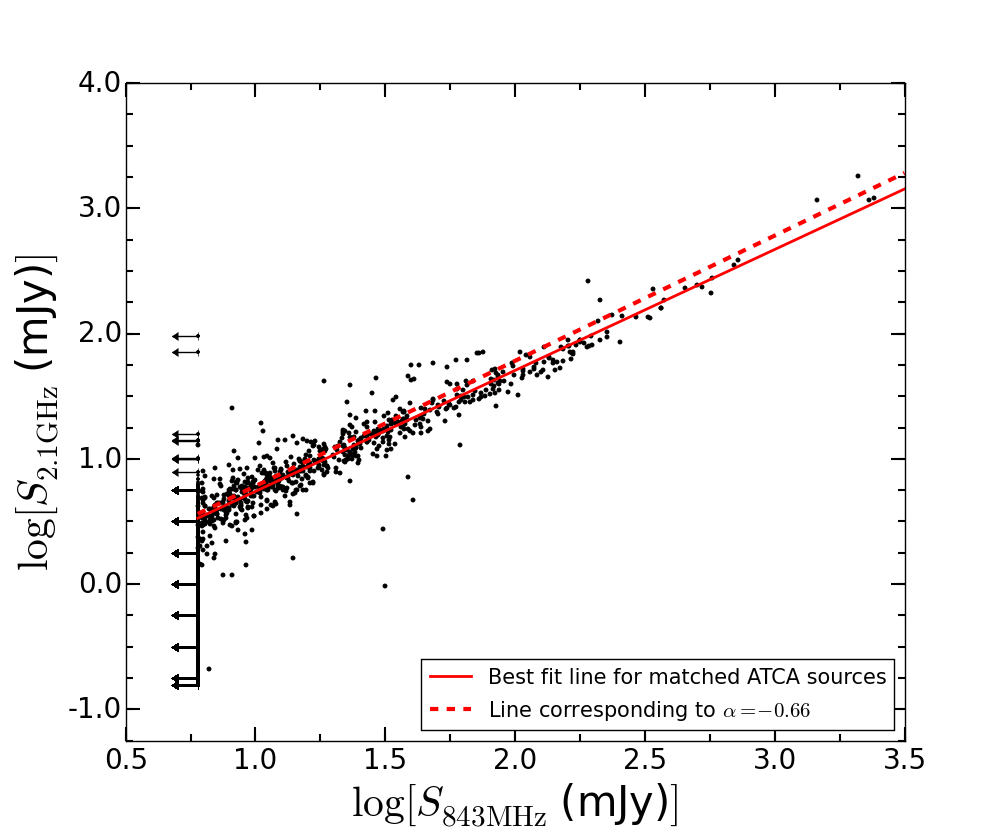}
    \caption{Log($S_{\rm{2.1GHz}}$) vs log($S_{\rm{843MHz}}$) for all ATCA XXL-S sources.  Those not matched to a SUMSS source were assigned a SUMSS upper limit of $S_{\rm{843MHz}}$ < 6 mJy.  The latter sources are clearly seen as the vertical line of data points with arrows at log($S_{\rm{843MHz}}$) $\approx$ 0.78 (arrows were plotted at regular intervals below log($S_{\rm{2.1GHz}}) \approx 0.9$ to avoid overcrowding the plot).  The solid red line is the best fit line for all ATCA sources matched to a SUMSS source, corresponding to $\alpha$~=~-0.81.  At log($S_{\rm{2.1GHz}}) \approx$ 0.48 ($S_{\rm{2.1GHz}} \approx$ 3 mJy), the spectral indices start to become highly censored because of the $S_{\rm{843MHz}}$~<~6 mJy limit.  The dashed red line is the line corresponding to $\alpha$~=~-0.66, the median spectral index of the ATCA sources with $S_{\rm{2.1GHz}}$~<~5 mJy (according to a survival analysis).  The difference in slope between the two lines reflects the bias toward a steeper median spectral index for ATCA sources matched to SUMSS sources.}
    \label{fig:S_XXL-S_vs_S_SUMSS}
\end{figure}

\begin{figure}
	\includegraphics[width=\columnwidth]{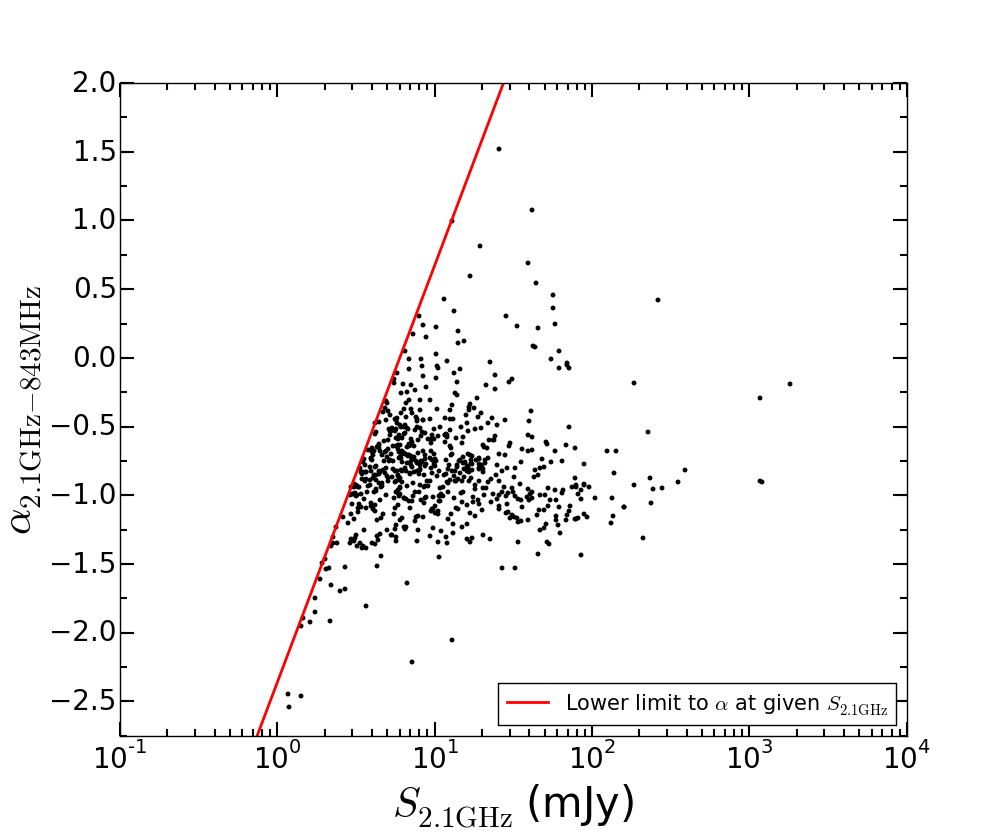}
    \caption{Spectral index vs $S_{\rm{2.1GHz}}$ for SUMSS sources cross-matched to one or more ATCA XXL-S sources.  The red line represents the lower limit to the spectral indices for matched ATCA sources with a given $S_{\rm{2.1GHz}}$.  It clearly demonstrates that sources with flatter spectral indices are missed at the faintest flux density levels ($S_{\rm{2.1GHz}}$~$\lesssim$~3 mJy.)}
    \label{fig:alpha_vs_S_XXL-S}
\end{figure}

However, selecting only SUMSS sources with at least one ATCA XXL-S match causes a bias in the spectral index distribution.  This is because the 5$\sigma$ SUMSS flux limit of 6 mJy is not sensitive enough to detect the bulk of ATCA XXL-S sources.  Therefore, at low 2.1 GHz flux densities, there is a bias toward steeper, more negative spectral indices (see \citealp{tasse2007} for a similar example).  Figure \ref{fig:S_XXL-S_vs_S_SUMSS}, which shows log($S_{\rm{2.1GHz}}$) vs log($S_{\rm{843MHz}}$) for all ATCA XXL-S sources, demonstrates this.  The solid red line is the best fit line for all ATCA sources matched to a SUMSS source, which corresponds to a spectral index of $\alpha = -0.81$ at the median log($S_{\rm{843MHz}}$) $\approx$ 1.21.  At $S_{\rm{843MHz}}$~=~6 mJy, the best fit line returns $S_{\rm{2.1GHz}} \approx$ 3 mJy.  For $S_{\rm{2.1GHz}}$ > 3 mJy, $\sim$16\% of the ATCA sources are not matched to a SUMSS source.  Below this $S_{\rm{2.1GHz}}$, this percentage increases rapidly, and the ATCA sources that are matched to a SUMSS source have highly censored spectral indices that are biased toward more negative values due to the SUMSS flux density limit.  For comparison, a survival analysis \citep{feigelson1985} was performed for the ATCA sources with $S_{\rm{2.1GHz}}$ < 5 mJy, which gave a median spectral index of $\alpha = -0.66^{+0.18}_{-0.07}$.  In Figure \ref{fig:S_XXL-S_vs_S_SUMSS}, the dashed red line corresponds to this value of $\alpha$.  The difference in slopes between these two lines illustrates that selecting only matched sources at all flux densities gives a biased representation of the spectral index distribution.  Figure \ref{fig:alpha_vs_S_XXL-S}, which is a plot of spectral index vs $S_{\rm{2.1GHz}}$ only for SUMSS sources cross-matched to ATCA XXL-S sources, also gives evidence for the bias.  For 3~mJy~<~$S_{\rm{2.1GHz}}$~<~10~mJy, the median spectral index starts to flatten.  The fainter ATCA sources have, on average, much steeper spectral indices than the brighter ones: the median spectral index for the 36 sources with $S_{\rm{2.1GHz}}$ < 3 mJy is $-1.53$, whereas the median spectral index for the 663 sources with $S_{\rm{2.1GHz}}$ > 3 mJy is $-0.81$.  This means that the sources with flatter spectral indices are missed below $S_{\rm{2.1GHz}}$ $\lesssim$ 3 mJy.

In order to avoid this censoring effect, ATCA sources with $S_{\rm{2.1GHz}}$ > 5 mJy were chosen to construct the spectral index distribution.  There are 582 ATCA sources above this flux limit, and only 5\% of them are not matched to a SUMSS source.  For these unmatched sources, a SUMSS upper limit of $S_{\rm{843MHz}}$ < 6 mJy was assigned, and hence the spectral indices of these sources represent their lower limits.  The median spectral index for the 553 matched ATCA sources with $S_{\rm{2.1GHz}}$~>~5 mJy is $\alpha$ = $-0.79$, which is significantly different to the median spectral index for the 29 unmatched ATCA sources with $S_{\rm{2.1GHz}}$ > 5 mJy, which is $\alpha \geq$ 0.07.

However, this represents a small percentage of the total number of ATCA XXL-S sources (86\% of the ATCA sample is not detected in SUMSS).  Therefore, a survival analysis incorporating every spectral index measurement and lower limit from all 6287 ATCA XXL-S sources was performed, resulting in a median value of $\alpha$ = $-0.75 \pm 0.03$.  This value is consistent with the median spectral index for ATCA sources with $S_{\rm{2.1GHz}}$ > 5 mJy ($\alpha$ = $-0.79$), but inconsistent with that for ATCA sources with $S_{\rm{2.1GHz}}$ < 5 mJy ($\alpha$ = $-0.66$).  This is a reflection of both the fact that survival analysis assumes the censored sources follow the distribution of the detected sources and of a probable systematic variation in spectral indices at fainter flux densities (e.g. \citealp{franzen2014}).  Hence, $\alpha$ = $-0.75$ may or may not be the actual median spectral index for the whole ATCA XXL-S sample. Nevertheless, this value is consistent with the typical $\alpha$ for AGN of $-0.7$ to $-0.8$ \citep{kimball2008}.

\section{Radio source counts}
\label{sec:source_counts}

The differential radio source counts were constructed using the source catalogue.  The flux density that was used for each source depended on whether or not it was resolved: $S_{\rm{int}}$ was used if the source was resolved, and $S_{\rm{p}}$ was used if it was unresolved.  The results are summarised in Table~\ref{tab:source_counts}.  Each line in the table shows the flux density interval $\Delta S$, the median flux density $S_{\rm{med}}$ of the sources in the flux density interval, the number of sources detected $N$, the number of sources after the completeness and flux density boosting corrections have been applied $N_{\rm{corr}}$, the corrected differential source counts $dN_{\rm{corr}}/dS$, and the corrected normalised source counts ($dN_{\rm{corr}}/dS)/S^{-2.5}$ with uncertainties.  The counts were normalised to the survey area (23.32 deg$^2 \approx 7.1 \times 10^{-3}$ sr) and the number expected in each flux density bin from the standard Euclidean count so that they could be compared to other surveys.  The flux density boosting corrections were applied to each source fainter than 0.381 mJy as $S_{\rm{corr}}=S/S_{\rm{frac}}$, where $S_{\rm{frac}}$ is the factor by which a given flux density $S$ is boosted, interpolated between the data points represented by the solid red line in Figure \ref{fig:s_frac_plot}.  The source count uncertainty for each flux density bin is defined as
\begin{equation}
\sigma_{\rm{sc}} = \sqrt{\sigma_{\rm{P}}^2 + \sigma_{\rm{c}}^2 + \sigma_{\rm{fb}}^2},
\end{equation}
where $\sigma_{\rm{P}}$ is the Poissonian uncertainty of $(\sqrt{N}/N) (dN/dS)/S^{-2.5}$, $\sigma_{\rm{c}}$ is the completeness correction uncertainty (the error bars in Figure \ref{fig:completeness_plot}) and $\sigma_{\rm{fb}}$ is the flux boosting correction uncertainty (the dashed red lines in Figure \ref{fig:s_frac_plot}).  Sources brighter than 0.381 mJy have no need for a flux density boosting correction, and therefore their $\sigma_{\rm{fb}}=0$.  Sources brighter than 0.873 mJy are in flux density bins that are 100\% complete, and therefore their $\sigma_{\rm{c}}=0$. 

\begin{figure*}
	\includegraphics[width=\textwidth]{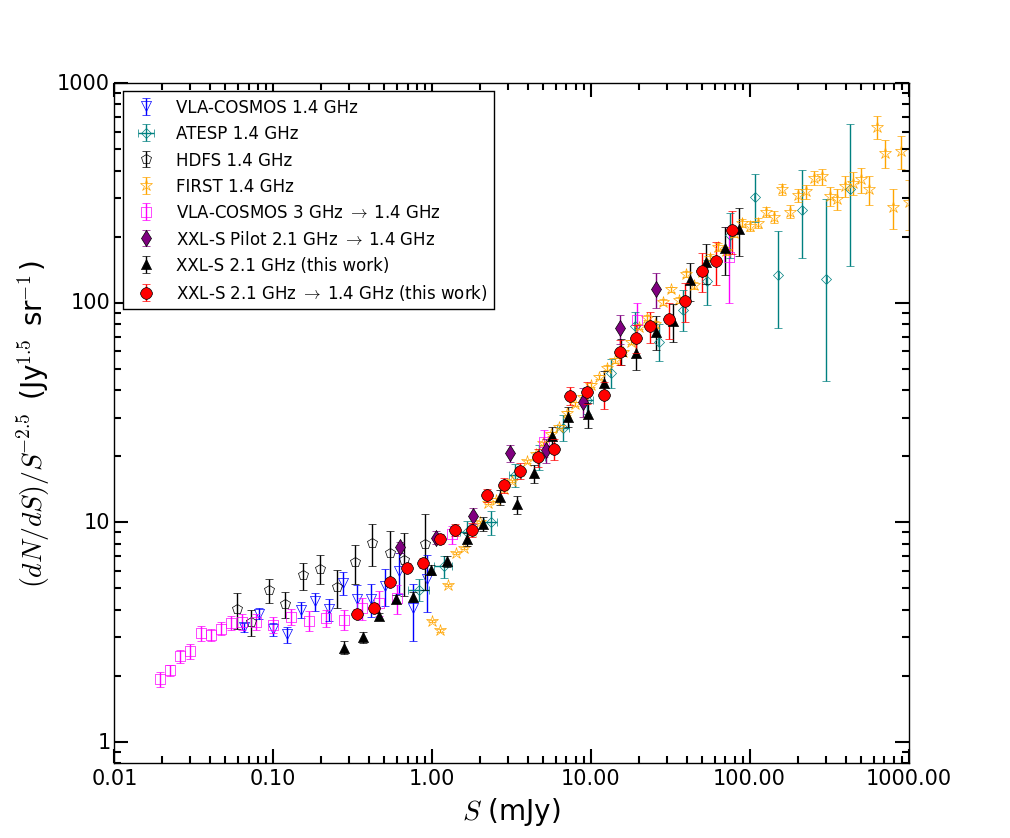}
    \caption{2.1 GHz source counts for XXL-S (black triangles).  The 1.4 GHz source counts, obtained by using a spectral index of $\alpha$ = $-$0.75 for each source, are shown as red circles for comparison to other 1.4 GHz surveys: VLA-COSMOS 1.4 GHz \citep{bondi2008}, ATESP \citep{prandoni2001}, HDFS \citep{huynh2005}, FIRST \citep{white1997}, VLA-COSMOS 3 GHz converted into 1.4 GHz \citep{smolcic2017_submitted}, and the pilot XXL-S survey converted into 1.4 GHz (XXL Paper XI).}
    \label{fig:source_counts_new_alpha}
\end{figure*}

Figure \ref{fig:source_counts_new_alpha} shows the 2.1 GHz source counts for XXL-S (black triangles) as well as the counts derived from converting the source flux densities into 1.4 GHz flux densities, using either the spectral indices described in Section \ref{sec:spectral_indices} or the median spectral index of $\alpha$~=~$-$0.75 (if the spectral index was unavailable or if the XXL-S source shared a SUMSS match with other XXL-S sources).  The minimum 1.4 GHz flux density plotted was determined by calculating the extrapolated 1.4 GHz flux density of the minimum 2.1 GHz flux density, again assuming a spectral index of $\alpha$ = $-$0.75.  This value was 0.302 mJy.  Adjusting the assumed spectral index between the values of $-$0.66 and $-$0.79 had an insignificant effect on the source counts.  The counts from other surveys are displayed in the figure for comparison.  The plot indicates that the XXL-S source counts extrapolated to 1.4 GHz are mostly in excellent agreement with other 1.4 GHz surveys. However, the counts start to flatten at a slightly higher flux density ($\sim$2 mJy) than the other surveys.  This is probably due to variation in the spectral indices as a function of flux density.

\begin{table*}
\centering
\caption{2.1 GHz source counts for XXL-S.  The columns are: flux density interval $\Delta S$, median flux density $S_{\rm{med}}$ of the sources in $\Delta S$, number of sources detected $N$, number of sources after the completeness and flux density boosting corrections have been applied $N_{\rm{corr}}$, corrected differential source counts $dN_{\rm{corr}}/dS$, and corrected normalised source counts ($dN_{\rm{corr}}/dS)/S^{-2.5}$ with uncertainties.}
\begin{tabular}{c r r r c c}
$\Delta S$ & \multicolumn{1}{c}{$S_{\rm{med}}$} & \multicolumn{1}{c}{$N$} & \multicolumn{1}{c}{$N_{\rm{corr}}$} & $dN_{\rm{corr}}/dS$ & ($dN_{\rm{corr}}/dS)/S^{-2.5}$\\
(mJy) & \multicolumn{1}{c}{(mJy)} & & & (Jy$^{-1}$ sr$^{-1}$) & (Jy$^{1.5}$ sr$^{-1}$)\\
\hline
\hline
0.250-0.321 & 0.282 & 859 & 1022 & 2.14e+09 & 2.85 $\pm$ 0.18\\
0.321-0.412 & 0.362 & 760 & 762 & 1.35e+09 & 3.35 $\pm$ 0.18\\
0.412-0.529 & 0.462 & 638 & 677 & 8.15e+08 & 3.73 $\pm$ 0.15\\
0.529-0.679 & 0.597 & 533 & 550 & 5.16e+08 & 4.49 $\pm$ 0.19\\
0.679-0.871 & 0.763 & 383 & 390 & 2.85e+08 & 4.59 $\pm$ 0.24\\
0.871-1.118 & 0.985 & 350 & 350 & 1.99e+08 & 6.06 $\pm$ 0.33\\
1.118-1.435 & 1.246 & 273 & 273 & 1.21e+08 & 6.64 $\pm$ 0.40\\
1.435-1.842 & 1.645 & 221 & 221 & 7.64e+07 & 8.40 $\pm$ 0.56\\
1.842-2.364 & 2.078 & 185 & 185 & 4.99e+07 & 9.81 $\pm$ 0.72\\
2.364-3.035 & 2.688 & 165 & 165 & 3.46e+07 & 12.98 $\pm$ 1.01\\
3.035-3.895 & 3.406 & 109 & 109 & 1.78e+07 & 12.07 $\pm$ 1.16\\
3.895-5.000 & 4.365 & 104 & 104 & 1.33e+07 & 16.69 $\pm$ 1.64\\
5.000-6.418 & 5.695 & 102 & 102 & 1.01e+07 & 24.78 $\pm$ 2.45\\
6.418-8.238 & 7.195 & 89 & 89 & 6.88e+06 & 30.23 $\pm$ 3.20\\
8.238-10.574 & 9.532 & 58 & 58 & 3.50e+06 & 31.01 $\pm$ 4.07\\
10.574-13.572 & 12.095 & 57 & 57 & 2.68e+06 & 43.06 $\pm$ 5.70\\
13.572-17.421 & 15.397 & 56 & 56 & 2.05e+06 & 60.26 $\pm$ 8.05\\
17.421-22.361 & 19.278 & 40 & 40 & 1.14e+06 & 58.82 $\pm$ 9.30\\
22.361-28.701 & 25.517 & 32 & 32 & 7.10e+05 & 73.89 $\pm$ 13.06\\
28.701-36.840 & 32.531 & 25 & 25 & 4.32e+05 & 82.53 $\pm$ 16.51\\
36.840-47.287 & 42.031 & 26 & 26 & 3.50e+05 & 126.89 $\pm$ 24.88\\
47.287-60.696 & 52.593 & 23 & 23 & 2.41e+05 & 153.17 $\pm$ 31.94\\
60.696-77.908 & 69.515 & 17 & 17 & 1.39e+05 & 177.15 $\pm$ 42.96\\
77.908-100.000 & 85.291 & 16 & 16 & 1.02e+05 & 216.60 $\pm$ 54.15\\
\end{tabular}
\label{tab:source_counts}
\end{table*}

\section{Summary}
\label{sec:conclusions}

The ATCA radio observations of the 25 deg$^2$ XXL-S field have been imaged to construct a 2.1 GHz radio mosaic, which achieved a resolution of $\sim$4.8$''$ and is sensitive down to $\sigma \approx 41$ $\mu$Jy/beam.  This is the largest area radio survey that has probed down to this flux density level.  The components with $S_{\rm{p}}>5\sigma$ in the mosaic were extracted with \textsc{blobcat}.  After removing spurious blobs and correcting each $S_{\rm{p}}$ for bandwidth smearing, the final radio component catalogue contained 6350 components, 1677 (26.4\%) of which were resolved.  Components with complex morphology were examined with DECam $z$-band imaging to determine whether each component group belonged to the same radio source.  One hundred and eleven components were joined together into 48 multiple-component sources, generating the source catalogue of 6287 sources.  Out of these, 1626 (25.9\%) are resolved.  The source catalogue was cross-matched to the SUMSS 843 MHz catalogue to examine the spectral indices of the SUMSS sources with $S_{\rm{843MHz}}>6$ mJy.  A survival analysis of all 6287 ATCA XXL-S sources found a median spectral index of $\alpha$ = $-$0.75.  Finally, the 1.4 GHz differential radio source counts corrected for completeness and flux density boosting were produced and overall are in excellent agreement with the other 1.4 GHz surveys presented.  Future work includes cross-matching the radio sources to the multiwavelength data available for XXL-S and constructing separate RLFs for the HERGs and LERGs in the field.

\section*{Acknowledgements}
XXL is an international project based around an XMM Very Large Programme surveying two 25 deg$^2$ extragalactic fields at a depth of $\sim$5 x 10$^{-15}$ erg cm$^{-2}$ s$^{-1}$ in the [0.5-2] keV band for point-like sources. The XXL website is http://irfu.cea.fr/xxl. Multi-band information and spectroscopic follow-up of the X-ray sources are obtained through a number of survey programmes, summarised at http://xxlmultiwave.pbworks.com/.  The Australia Telescope Compact Array is part of the Australia Telescope National Facility which is funded by the Australian Government for operation as a National Facility managed by CSIRO.  A.B. acknowledges the University of Western Australia (UWA) for funding support from a University Postgraduate Award PhD scholarship and the International Centre for Radio Astronomy Research (ICRAR) for additional support.  A.B. also thanks Tom Franzen for assistance on the peeling and \textsc{clean} bias simulation.  M.H. thanks UWA for a Research Collaboration Award (RCA) grant to collaborate with the COSMOS and XXL team at the University of Zagreb.  V.S. and N.B. acknowledge funding from the European Union's Seventh Framework program under grant agreement 333654 (CIG, `AGN feedback'). V.S., J.D., and M.N. acknowledge funding from the European Union's Seventh Framework program under grant agreement 337595 (ERC Starting Grant, `CoSMass'). V.S. acknowledges support from the ICRAR Visiting Fellowship For Senior Women In Astronomy 2015.  All the authors thank the referee for her helpful insights and suggestions for improving the paper, as well as Mark Birkinshaw, Huub R\"{o}ttgering, Chiara Ferrari, Sean McGee, Somak Raychaudhury, Florian Pacaud, Andrea Comastri, Malcolm Bremer, Nick Tothill, Ray Norris, Marco Bondi, and Paolo Ciliegi for their contribution to the ATCA proposal (project C2627).

\clearpage

\begin{landscape}

\begin{table}
\caption{Example entries from the 2.1 GHz component catalogue for XXL-S (XXL\_ATCA\_16A).  Components 337-1 and 337-2 are separate components from a de-blended blob.  Column descriptions can be found in Section \ref{sec:final_cats}.}

\begin{adjustbox}{width=24.7cm}
\begin{tabular}{l l c c c c c c c c c c c c c c c c}
\multicolumn{1}{c}{IAU name} & ID & RA & Dec & $\sigma_{\rm{RA}}$ & $\sigma_{\rm{Dec}}$ & $\sigma_{\rm{rms}}$ & $S_{\rm{p}}$ & $\sigma_{S_{\rm{p}}}$ & $S_{\rm{int}}$ & $\sigma_{S_{\rm{int}}}$ & $S/N$ & Res & MC & $\Theta$ & $\sigma_{\Theta}$ & $\nu_{\rm{eff}}$ & Complex\\
 & & (deg) & (deg) & ($''$) & ($''$) & (mJy/beam) & (mJy/beam) & (mJy/beam) & (mJy) & (mJy) & & & & ($''$) & ($''$) & (GHz) & \\
\hline
\hline
2XXL-ATCA J232805.5-554110A &90    &352.030820 &-55.685561 &0.02 &0.02 &0.0744&   27.5472 & 1.4352 &  42.6658 & 2.1346 & 370.3 &1 &1 & 3.529 &0.356 &1.7991 &1\\
2XXL-ATCA J233106.6-530426  &231   &352.777506 &-53.073905 &0.02 &0.03 &0.0398&    6.7734 & 0.3546 &   6.9134 & 0.3480 & 170.2 &0 &0 & 0.000 &0.000 &1.8139 &0\\
2XXL-ATCA J234438.3-534725  &337-1 &356.159769 &-53.790543 &0.05 &0.06 &0.0354&    4.1764 & 0.1038 &   4.3739 & 0.1624 & 118.0 &0 &0 & 0.000 &0.000 &1.8051 &0\\
2XXL-ATCA J234437.0-534731  &337-2 &356.154480 &-53.792200 &0.28 &0.27 &0.0354&    0.8420 & 0.0645 &   2.3265 & 0.2716 &  23.8 &1 &0 & 6.325 &0.418 &1.8048 &0\\
2XXL-ATCA J234550.2-560019  &406   &356.459465 &-56.005457 &0.03 &0.04 &0.0359&    3.4560 & 0.1834 &   3.8609 & 0.1964 &  96.3 &1 &0 & 1.631 &0.566 &1.7952 &1\\
2XXL-ATCA J234004.8-544438A &1000  &355.017937 &-54.739532 &0.09 &0.12 &0.0445&    1.5101 & 0.0903 &   3.8678 & 0.1984 &  33.9 &1 &1 & 5.952 &0.381 &1.8105 &1\\
2XXL-ATCA J234756.3-542722  &1370  &356.984903 &-54.456388 &0.13 &0.17 &0.0344&    0.8276 & 0.0551 &   0.8547 & 0.0549 &  24.1 &0 &0 & 0.000 &0.000 &1.7988 &0\\
2XXL-ATCA J233204.8-524636  &2040  &353.020155 &-52.776896 &0.21 &0.27 &0.0544&    0.8136 & 0.0689 &   0.9961 & 0.0737 &  15.0 &1 &0 & 2.256 &0.690 &1.8054 &0\\
2XXL-ATCA J232707.9-554233  &3011  &351.783035 &-55.709210 &0.32 &0.41 &0.0463&    0.4504 & 0.0519 &   0.4107 & 0.0506 &   9.7 &0 &0 & 0.000 &0.000 &1.8014 &0\\
2XXL-ATCA J234222.9-550243  &4665  &355.595670 &-55.045394 &0.49 &0.62 &0.0343&    0.2195 & 0.0361 &   0.2517 & 0.0365 &   6.4 &0 &0 & 0.000 &0.000 &1.8105 &0\\
2XXL-ATCA J234310.8-555250C &5099  &355.795991 &-55.881405 &0.51 &0.66 &0.1380&    0.8326 & 0.1442 &   0.8085 & 0.1433 &   6.0 &0 &1 & 0.000 &0.000 &1.7969 &0\\
2XXL-ATCA J234956.9-544228  &6271  &357.487167 &-54.708013 &0.59 &0.76 &0.0388&    0.2038 & 0.0402 &   0.1711 & 0.0397 &   5.3 &0 &0 & 0.000 &0.000 &1.8159 &0\\

\end{tabular}
\end{adjustbox}
\label{tab:comp_catalogue}
\end{table}

\begin{table}
\caption{Example entries from the 2.1 GHz source catalogue for XXL-S (XXL\_ATCA\_16B).  These entries are all multiple-component sources (the single-component sources have identical entries in the component catalogue).  Column descriptions can be found in Section \ref{sec:final_cats}.}

\begin{adjustbox}{width=24.7cm}
\begin{tabular}{l l c c c c c c c c c c c c c c c c}
\multicolumn{1}{c}{IAU name} & ID & RA & Dec & $\sigma_{\rm{RA}}$ & $\sigma_{\rm{Dec}}$ & $\sigma_{\rm{rms}}$ & $S_{\rm{p}}$ & $\sigma_{S_{\rm{p}}}$ & $S_{\rm{int}}$ & $\sigma_{S_{\rm{int}}}$ & $S/N$ & Res & MC & $\Theta$ & $\sigma_{\Theta}$ & $\nu_{\rm{eff}}$ & Complex\\
 & & (deg) & (deg) & ($''$) & ($''$) & (mJy/beam) & (mJy/beam) & (mJy/beam) & (mJy) & (mJy) & & & & ($''$) & ($''$) & (GHz) & \\
\hline
\hline
2XXL-ATCA J232805.5-554110 & 90\_362 & 352.023228 & -55.686247 & 0.03 & 0.04 & 0.0759 & -99 & -99 & 61.4915 & 2.3341 & -99 & 1 & 1 & -99 & -99 & 1.7992 & 1\\
2XXL-ATCA J232624.7-524209 & 131\_300 & 351.603286 & -52.702566 & 0.03 & 0.04 & 0.0828 & -99 & -99 & 59.1999 & 2.1615  & -99 & 1  & 1  & -99 & -99 & 1.7984  & 1 \\
2XXL-ATCA J234323.7-560342   & 223\_265         & 355.849028  & -56.061678  & 0.03  & 0.04  & 0.0677 & -99 & -99 &   44.5605  &  1.6255  & -99 & 1  & 1  & -99 & -99 & 1.7953  & 1 \\
2XXL-ATCA J232904.7-563453   & 430\_528\_2116    & 352.269836  & -56.581437  & 0.23  & 0.29  & 0.0652 & -99 & -99 &   13.7939  &  0.4646  & -99 & 1  & 1  & -99 & -99 & 1.8564  & 1 \\
2XXL-ATCA J234705.2-534138   & 638\_2971        & 356.772001  & -53.694144  & 0.37  & 0.47  & 0.1077 & -99 & -99 &   41.4072  &  1.5341  & -99 & 1  & 1  & -99 & -99 & 1.8018  & 1 \\
2XXL-ATCA J233915.3-535746   & 844\_1800        & 354.814060  & -53.962858  & 0.20  & 0.25  & 0.0449 & -99 & -99 &    8.1056  &  0.3152  & -99 & 1  & 1  & -99 & -99 & 1.8098  & 1 \\
2XXL-ATCA J233102.2-533220   & 919\_1284        & 352.759451  & -53.538913  & 0.15  & 0.19  & 0.0457 & -99 & -99 &    8.3143  &  0.3025  & -99 & 1  & 1  & -99 & -99 & 1.8033  & 1 \\
2XXL-ATCA J234837.1-552621   & 1226\_2070\_2269  & 357.154638  & -55.439243  & 0.34  & 0.44  & 0.0413 & -99 & -99 &    6.4243  &  0.2091  & -99 & 1  & 1  & -99 & -99 & 1.8048  & 1 \\
2XXL-ATCA J232828.2-530332   & 1364\_2142       & 352.117685  & -53.059091  & 0.26  & 0.33  & 0.0509 & -99 & -99 &    6.1075  &  0.2287  & -99 & 1  & 1  & -99 & -99 & 1.8177  & 1 \\
2XXL-ATCA J233947.3-544926   & 1758\_1844\_4147  & 354.947330  & -54.824113  & 0.51  & 0.65  & 0.0479 & -99 & -99 &    8.4038  &  0.2995  & -99 & 1  & 1  & -99 & -99 & 1.8380  & 1 \\
2XXL-ATCA J234310.8-555250   & 2068\_2724\_5099  & 355.795241  & -55.880827  & 0.65  & 0.83  & 0.1375 & -99 & -99 &   45.0891  &  1.5790  & -99 & 1  & 1  & -99 & -99 & 1.7969  & 1 \\
2XXL-ATCA J232857.5-533724   & 3024\_3157       & 352.239790  & -53.623591  & 0.48  & 0.62  & 0.0574 & -99 & -99 &    4.4161  &  0.1854  & -99 & 1  & 1  & -99 & -99 & 1.7613  & 1 \\

\end{tabular}
\end{adjustbox}
\label{tab:source_catalogue}
\end{table}

\end{landscape}



\clearpage

\bibliographystyle{aa}
\bibliography{paper_1_aa}




\clearpage

\appendix

\section{sub-band imaging analysis}
\label{sec:sb}

\begin{table}
\caption{Central frequencies, robust values and image radii for the seven 256-MHz-wide sub-bands.}
\begin{tabular}{c c c}
Central frequency & Robust parameter & Image radius\\
(MHz) & & (arcmin)\\
\hline
\hline
1460 & -2.0 & 31.6\\
1716 & -2.0 & 27.8\\
1972 & -0.3 & 25.0\\
2228 & 0.0 & 22.5\\
2484 & 0.15 & 20.1\\
2740 & 0.3 & 18.1\\
2996 & 0.4 & 16.5\\
\hline
\end{tabular}
\label{tab:sb_freq_robust}
\end{table}

As mentioned in Section \ref{sec:imaging}, the primary beam response, synthesised beam size, and flux densities of most sources vary significantly from 1.1 to 3.1 GHz.  Therefore, an improvement in the quality of the data was sought by using \texttt{UVSPLIT} to divide the 2 GHz bandwidth into eight 256-MHz-wide sub-bands from 1076-3124 MHz and combining those sub-bands together.  This approach has been used with 2-4 GHz VLA data \citep{novak2015, smolcic2015a} and was also adopted for the pilot data (XXL Paper XI).  The 1076-1332 MHz sub-band was thrown out due to high RFI contamination, so seven sub-bands were used in the analysis.  The steps to image every pointing in each of these seven sub-bands were identical to those for the full-band, with the following exceptions: different robust parameters, different image dimensions corresponding to the radii at which the primary beam responses were $\sim$8\% for the given central frequencies, a \texttt{SELFCAL} interval of 0.33 minutes (which was found to minimise the image artefacts around bright sources), and a different \texttt{CONVOL} beam size of 6.7$''$ $\times$ 5.5$''$ with a position angle of 3.09$^{\circ}$.  Different robust parameters were necessary for making the synthesised beam sizes for each sub-band as similar as possible before convolving the images. Table \ref{tab:sb_freq_robust} lists the central frequencies, robust parameters and the radii the individual pointings were imaged out to for each sub-band.

\begin{figure}
	\includegraphics[width=\columnwidth]{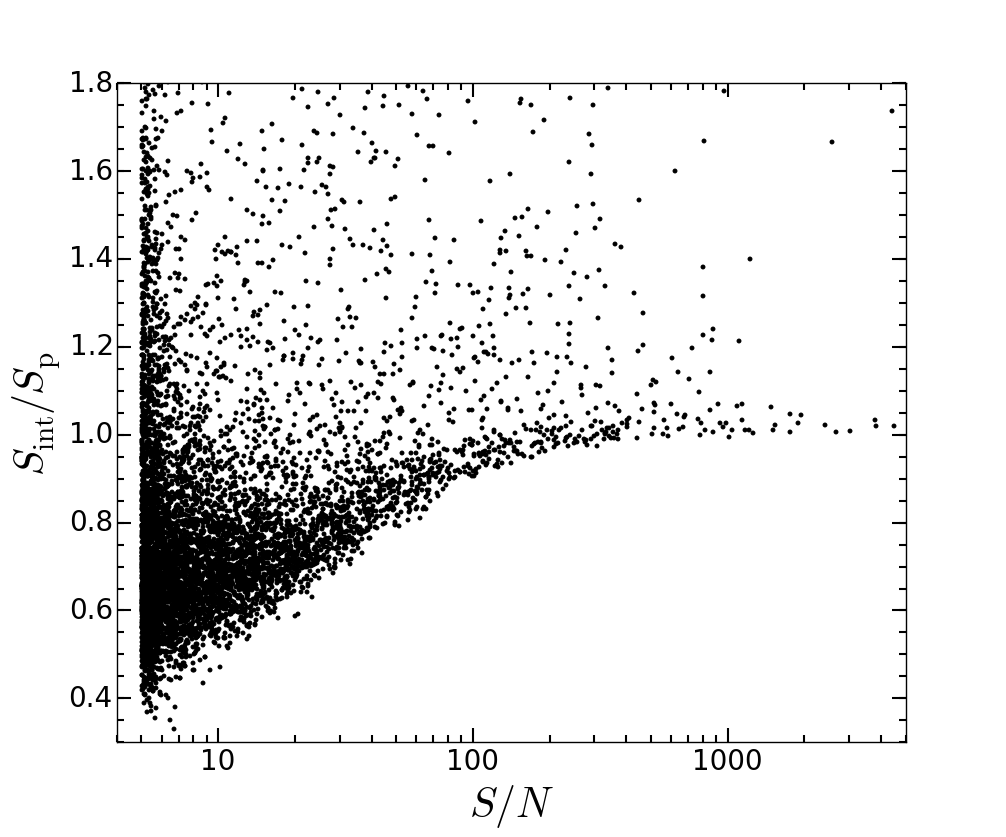}
    \caption{$S_{\rm{int}}/S_{\rm{p}}$ vs $S/N$ plot for the sub-band combined mosaic.  There is a clear systematic in $S_{\rm{int}}/S_{\rm{p}}$ that was not possible to eliminate.  This trend also appeared in the individual sub-band mosaics, indicating that fractional bandwidth was not the cause.}
    \label{fig:sb-comb_Sint_Sp_SNR_plot}
\end{figure}

Once each pointing was imaged in each sub-band, \texttt{LINMOS} was used to combine and mosaic all the images together.  An $S_{\rm{int}}/S_{\rm{p}}$ vs $S/N$ plot was made for this sub-band combined mosaic, which is shown in Figure \ref{fig:sb-comb_Sint_Sp_SNR_plot}.  This plot shows a clear downward trend in $S_{\rm{int}}/S_{\rm{p}}$ with decreasing $S/N$.  This effect showed up in the individual sub-band mosaics as well (not just the one with all the sub-bands combined), indicating that the fractional bandwidth was not the cause of the problem.  If this trend was not eliminated, the integrated flux densities would have a very large negative bias and it would not be possible to reliably determine which sources are resolved and which ones are not.

Therefore, a number of tests were executed in an attempt to remove the trend.  The tests were:

\begin{enumerate}
\item Tapering the visibilities in each sub-band $uv$ data file in the \texttt{INVERT} step of the imaging in order to achieve similar beam sizes.
\item Imaging the full-band data first, splitting the full-band $uv$ data into the seven sub-bands with \texttt{UVSPLIT}, and then cleaning each sub-band down to 6$\sigma$ using the parameters listed in Table \ref{tab:sb_freq_robust}.
\item Using the same image radius for all sub-bands.
\item Restoring the clean component models with the average beam size for each sub-band.
\end{enumerate}
All of these tests had some effect on the shape of the $S_{\rm{int}}/S_{\rm{p}}$ vs $S/N$ plot, but none were able to completely eliminate the trend in a way that reproduced the $S_{\rm{int}}/S_{\rm{p}}$ vs $S/N$ plot for the full-band.  Test 4 came the closest to flattening the trend, but implementing it would have resulted in the cleaned sources having a different resolution to the noise, rendering source extraction unreliable.

Furthermore, the sub-band images have larger rms noise values due to the smaller number of channels (i.e. less $uv$ coverage), and therefore it is not possible to clean as deeply in the sub-band as it is in the full-band.  This results in fewer sources being included in the clean component model for the individual images.  When the images are combined into the mosaic, the noise floor is decreased, allowing some uncleaned sources (i.e. sources not included in the clean model for individual pointings) to rise above the detection threshold of the source extractor.  These uncleaned sources have been convolved with the dirty beam, not the clean beam, and yet remain in the final mosaic image.  For a given flux density, therefore, there would be discrepancies between the sizes of uncleaned sources and their corresponding sizes if they had been cleaned, and the discrepancies would increase with decreasing $S/N$.  This effect is the probable cause of the trend in the sub-band combined $S_{\rm{int}}/S_{\rm{p}}$ vs $S/N$ plot.  In addition, using the sub-band combined mosaic would have decreased the overall sensitivity of the source extraction process.  The peak rms noise value contained in the sub-band combined mosaic ($\sigma \approx 55$ $\mu$Jy) is higher than the full-band's by about 50\%.  Consequently, the full-band mosaic was chosen for source extraction because there was no systematic in the flux densities of the sources and because of its greater sensitivity to cleaned sources near the 5$\sigma$ detection threshold.

\label{lastpage}
\end{document}